\documentclass{article}
\usepackage[utf8]{inputenc}
\usepackage[gen]{eurosym}
\usepackage{listings}   % pachetul principal
\usepackage{courier}    % font monospatiat frumos
\usepackage{xcolor}     % culori (optional)

\usepackage{color}
\usepackage{amsmath}
\usepackage{graphicx}
\usepackage{footnote}
\usepackage{subscript}
\usepackage{amsfonts,color}
\usepackage{amsmath,amssymb}
\usepackage{mathtools}
\usepackage{epsfig}
\usepackage{amssymb}
\usepackage{mathtools}
\usepackage{amsthm,wasysym}
\usepackage[english]{babel}
\usepackage{cancel}
\usepackage{colonequals}
\makeatletter
\usepackage{float}
\usepackage{wrapfig}
\usepackage{nicefrac,enumitem}
\usepackage{hyperref}
\usepackage{fixltx2e}
\restylefloat{figure}

\makeatletter
\let\@fnsymbol\@arabic
\makeatother

\usepackage{bbm}
\newcommand{\id}{{\boldsymbol{\mathbbm{1}}}}

\newcommand{\tr}{{\rm tr}}
\newcommand{\dev}{{\rm dev}}
\newcommand{\sym}{{\rm sym}}
\newcommand{\skw}{{\rm skew}}

\newcommand{\norm}[1]{\|#1\|}

\def\dd{\displaystyle}

\newtheorem{theorem}{Theorem}[section]

\newtheorem{remark}[theorem]{Remark}
\newtheorem{proposition}[theorem]{Proposition}
\newtheorem{corollary}[theorem]{Corollary}

\newtheorem{definition}[theorem]{Definition}

\def\dd{\displaystyle}

\usepackage{multicol}
\usepackage[font=small]{caption}
\newcommand{\citet}[2][]{\citeauthor{#2} \cite[#1]{#2}}
\graphicspath{{gfx/}}
\RequirePackage{amsmath}	% provides "\DeclareMathOperator" command
\RequirePackage{mathtools}	% provides "\DeclarePairedDelimiter" command
\newcommand{\setvert}{\,|\,}

\newcommand{\col}{\colon}

\newcommand{\C}{\mathbb{C}}
\newcommand{\R}{\mathbb{R}}

\DeclareMathOperator{\GL}{GL}

\DeclareMathOperator{\SO}{SO}
\DeclareMathOperator{\OO}{O}

\DeclareMathOperator{\Sym}{Sym}

\newcommand{\Rmat}[2]{\mathbb{R}^{#1 ? #2}}
\newcommand{\Rnn}{\Rmat{n}{n}}

\newcommand{\GLp}{\GL^{\!+}}

\newcommand{\GLpn}{\GLp(n)}

\newcommand{\On}{\OO(n)}

\newcommand{\Symn}{\Sym(n)}

\DeclareMathOperator{\diag}{diag}

\DeclareMathOperator{\Cof}{Cof}

\newcommand{\intd}[1]{{\mathrm{d}#1}}

\newcommand{\dx}{\intd{x}}

\newcommand{\pdd}[3][]{\frac{\partial\ifx&#1&\else^{#1}\fi #2}{\partial #3}}

\DeclareMathOperator{\@macros@div}{div}
\renewcommand{\div}{\@macros@div}

\newcommand{\D}{\mathrm{D}}
\newcommand{\innerproduct}[1]{\langle #1 \rangle}

\newcommand{\iprod}{\innerproduct}

\newcommand{\tel}[1]{\frac{1}{#1}}

\providecommand{\availableaturl}[2][]{%
	available at \url{#2}%
}

\usepackage[babel,german=quotes]{csquotes}

\makeatletter%
\@ifpackageloaded{hyperref}{}{}%
\makeatother%

\usepackage{color}
\usepackage{amsmath}

\usepackage{footnote}
\usepackage{subscript}
\usepackage{amsfonts,color}
\usepackage{amsmath,amssymb}
\usepackage{amssymb}
\usepackage{mathtools}
\usepackage{amsthm,wasysym}
\usepackage{framed}
\usepackage{cancel}
\usepackage{graphicx}
\usepackage{caption}
\usepackage{subcaption}
\usepackage{rotating}

\usepackage[english]{babel}
\usepackage[utf8]{inputenc}
\makeatletter
\usepackage{float}
\usepackage{wrapfig}
\restylefloat{figure}

\makeatletter
\let\@fnsymbol\@arabic
\makeatother

\usepackage{bbm}

\def\dd{\displaystyle}

\def\barr{\begin{array}}
	\usepackage{lscape}

	\def\earr{\end{array}}
\def\bec#1{\begin{equation}\label{#1}}
\def\becn{\begin{equation*}}
\def\endec{\end{equation}}
\def\endecn{\end{equation*}}
\def\dd{\displaystyle}
\def\bfm#1{\mbox{\boldmat}}

\renewcommand{\dd}{\displaystyle}

\usepackage{enumitem}

\usepackage{pifont}% http://ctan.org/pkg/pifont

\newcommand{\imin}[1][\lambda]{m{\ifx&#1&\else(#1)\fi}}
\newcommand{\imax}[1][\lambda]{M{\ifx&#1&\else(#1)\fi}}
\newcommand{\iset}[1][\lambda]{J{\ifx&#1&\else(#1)\fi}}
\allowdisplaybreaks[1]

\usepackage{titletoc}

\begin{document}
\title{Constitutive properties  for  isotropic energies in ideal  nonlinear  elasticity for solid materials: numerical evidences for invertibility and monotonicity in different stress-strain pairs }
\author{
	Ionel-Dumitrel Ghiba\thanks{Ionel-Dumitrel Ghiba,   Alexandru Ioan Cuza University of Ia\c si, Department of Mathematics,  Blvd. Carol I, no. 11, 700506 Ia\c si,
		Romania;  Octav Mayer Institute of Mathematics of the
		Romanian Academy, Ia\c si Branch,  700505 Ia\c si, email:  dumitrel.ghiba@uaic.ro},   \qquad Robert J. Martin\thanks{Robert J. Martin,  Lehrstuhl f\"{u}r Nichtlineare Analysis und Modellierung, Fakult\"{a}t f\"{u}r Mathematik, Universit\"{a}t Duisburg-Essen, Thea-Leymann Str. 9, 45127 Essen, Germany, email: robert.martin@uni-due.de}, \qquad Marius Apetrii\thanks{Marius Apetrii,   Alexandru Ioan Cuza University of Ia\c si, Department of Mathematics,  Blvd. Carol I, no. 11, 700506 Ia\c si,
		Romania, email:  mapetrii@uaic.ro},\vspace{1.2mm}\\ \quad and Patrizio Neff\thanks{Patrizio Neff,  Head of Lehrstuhl f\"{u}r Nichtlineare Analysis und Modellierung, Fakult\"{a}t f\"{u}r Mathematik, Universit\"{a}t Duisburg-Essen,  Thea-Leymann Str. 9, 45127 Essen, Germany, email: patrizio.neff@uni-due.de} }
	
	\date{\it Dedicated to Professor Philippe G. Ciarlet with great admiration}

\maketitle

\begin{abstract}
		
As a service for the  solid mechanics community we  gather in this paper constitutive properties of a collective list of  isotropic elastic energies for compressible materials. Of interest to us are the invertibility and monotonicity of certain stress-strain pairs. The calculations are done numerically by our own evaluation algorithm and presented in a yes/no-table. Such an overview has been missing up to now. It is intended to expand the table with further energies as  time goes on and updates  will be found on arxive.
	
\end{abstract}
\setcounter{tocdepth}{1}

\tableofcontents

\section{Introduction}
The theory of elasticity has without doubt many applications. However, depending on the phenomena we intend to study, various elastic energies have been considered. Therefore, we reconsider the question of  Truesdell (1919-2000) formulated in 	 ``Das ungel\"oste Hauptproblem der endlichen Elastizit\"atstheorie''. Zeit. Angew. Math.  Mech. 36 (3-4):  97-103, 1956:
	\begin{center}		
{	\it	``Welches ist die Klasse der Funktionen {\rm [$W$]}, die als Form\"anderungsarbeitsdichte   eines vollkommen elastisches Stoffes dienen d\"urfen?''

		\noindent  ``What is the class of functions {\rm [$W$]}, which can serve as strain energy densities, for perfectly elastic materials?''
		
	 }
	\end{center}
	
	In trying to partly adress  this question, in the following
		 we aim to numerically exhibit certain properties of the below  (incomplete and biased list) of constitutive choices. We deliberately skip standard conditions like polyconvexity or Legendre-Hadamard-ellipticity which ensure a kind of material stability  for the mathematical  analysis. To the contrary, here we concentrate on constitutive issues believed to be more closely related to the underlying mechanics and physics. In this respect, we identify invertibility and monotonicity (``stress increases with strain'') as  most important. More precisely, we consider 3 stress tensors: the spatial Cauchy stress $\sigma$, the spatial Kirchhoff stress $\tau=\det\, F\cdot \sigma$ and the referential Biot-stress $T_{\rm Biot}=\D_U\widetilde{W}(U)$ (the nominal or engineering stress) and determine  local invertibility and monotonicity properties with respect to the stretch $V$ (or $U$) and the logarithmic stretch $\log V$ (or $\log U$). Both, {\bf local invertibility} and {\bf local monotonicity} are checked numerically only and we skip the attribute ``local" in the following (see the Appendix \ref{appendixA} for the definitions of all used quantities  from nonlinear hyperelasticity).
		
	Note that every isotropic and frame-invariant function\footnote{An elastic energy potential $W\col\GLp(n)\to\R$ is also often assumed to be \emph{objective} (or \emph{frame-indifferent}) as well as \emph{isotropic}, i.e.\ it satisfies
			\begin{equation}
				W(Q_1F\,Q_2) = W(F)
				\qquad\text{for all }\;F\in\GLp(n)
				\quad\text{and all }\;\; Q_1,Q_2\in\SO(n)\,,
			\end{equation}
			where $\SO(n)=\{X\in \mathbb{R}^{n\times n} \setvert Q^T Q=\id\,,\;\det Q=1\}$ denotes the special orthogonal group.} of $F$ is  expressible in the form
\begin{align}
	W(F) &=\widetilde{W}(U)=\overline{W}(C)=\widehat{W}(
	\log V)=g(\lambda_1, \lambda_2, \lambda_3),
\end{align}
where $\lambda_i=\lambda_i(\sqrt{F^TF})$ are the singular values of $F$ (or eigenvalues of $\sqrt{F^TF}$).
The functions $\widehat{W}$ and $ g$ are uniquely determined by $W$. We consider \begin{align}
\widehat{T}_{\rm Biot}(\log U)&:={T}_{\rm Biot}(U)\notag,\qquad \qquad\widehat{\sigma}(\log V):=\sigma(V),\qquad \qquad\widehat{\tau}(\log V):=\tau(V).
\end{align} Thus, 	we say that
	\begin{align}
	U\mapsto T_{\rm Biot}(U)\qquad  &\text{is invertible in}\qquad U  & \Longleftrightarrow & \qquad  \qquad \det{\rm D}_UT_{\rm Biot}(U)>0,\notag\\
	V\mapsto 	\tau(V)\qquad  &\text{is invertible in}\qquad V & \Longleftrightarrow& \qquad  \qquad \det{\rm D}_V\tau(V)>0,\notag\\
		V\mapsto \sigma(V)\qquad  &\text{is invertible in}\qquad V &\hspace*{-0.5cm}\Longleftrightarrow& \qquad  \qquad \det{\rm D}_V\sigma(V)>0,\\
	\log U\mapsto 	\widehat{T}_{\rm Biot}(\log U)\qquad  &\text{is invertible in}\qquad \log U & \Longleftrightarrow& \qquad  \qquad \det{\rm D}_{\log U}\widehat{T}_{\rm Biot}(\log U)>0,\notag\\
	\log V\mapsto \widehat{\tau}(\log V)\qquad  &\text{is invertible in}\qquad \log V & \Longleftrightarrow& \qquad  \qquad \det{\rm D}_{\log V}\widehat{\tau}(\log V)>0,\notag\\
	\log V\mapsto \widehat{\sigma}(\log V)\qquad  &\text{is invertible in}\qquad \log V & \Longleftrightarrow& \qquad  \qquad \det{\rm D}_{\log V}\widehat{\sigma}(\log V)>0.\notag
	\end{align}
	
	We also note that
	\begin{align}
		T_{\rm Biot}:={\rm D}_U[\widetilde{W}(U)],\qquad \qquad \text{while}\qquad \widehat{\tau}={\rm D}_{\log V}[\widehat{W}({\log V})],
		\end{align}
		due to Richter's formula \cite{NeffGhibaLankeit}.
		
	It is  clear that (local) invertibility with respect to $U$ (or $V$) is equivalent to local invertibility with respect to $\log U$ (or $\log V$) since ${\rm Sym}^{++}(3)\ni U\mapsto \log\, U\in {\rm Sym}(3)$ is itself an  invertible mapping. Similarly, we say that
	\begin{align}\label{1.5}
		T_{\rm Biot}\qquad  &\text{is monotone in}\qquad U  &\hspace*{-0.5cm}\Longleftrightarrow  &\qquad  \qquad\underbrace{{\rm \ D}_UT_{\rm Biot}(U)}_{\text{\rm  {\bf not} major symmetric}}&\hspace{-1.5cm}\in {\rm Sym}_4^{++}(6),\notag\\
		\tau\qquad  &\text{is monotone in}\qquad V &\hspace*{-0.5cm}\Longleftrightarrow&\qquad  \qquad\sym\,\underbrace{{\rm D}_V\tau(V)}_{\text{\rm  {\bf not} major symmetric}}&\hspace{-1.5cm}\in {\rm Sym}_4^{++}(6),\notag\\
		\sigma\qquad  &\text{is monotone in}\qquad V &\hspace*{-0.5cm}\Longleftrightarrow &\qquad  \qquad\sym\,\underbrace{ {\rm D}_V\sigma(V)}_{\text{\rm  {\bf not} major symmetric}}&\hspace{-1.5cm}\in {\rm Sym}_4^{++}(6),\\
		\widehat{T}_{\rm Biot}\qquad  &\text{is monotone in}\qquad \log U &\hspace*{-0.5cm}\Longleftrightarrow& \qquad  \qquad\sym\underbrace{ {\rm D}_{\log U}\widehat{T}_{\rm Biot}(\log U)}_{\text{\rm   {\bf not} major symmetric}}&\hspace{-1.5cm}\in {\rm Sym}_4^{++}(6),\notag\\
		\widehat{\tau}\qquad  &\text{is monotone in}\qquad \log V &\hspace*{-0.5cm}\Longleftrightarrow &\qquad  \qquad\underbrace{{\rm D}_{\log V}\widehat{\tau}(\log V)}_{\text{\rm major symmetric}}&\hspace{-1.5cm}\in {\rm Sym}_4^{++}(6),\notag\\
		\widehat{\sigma}\qquad  &\text{is monotone in}\qquad \log V &\hspace*{-0.5cm}\Longleftrightarrow&\qquad  \qquad\sym \underbrace{{\rm D}_{\log V}\widehat{\sigma}(\log V)}_{\text{\rm  {\bf not} major symmetric}}&\hspace{-1.5cm}\in {\rm Sym}_4^{++}(6).\notag
	\end{align}
	
	We observe  that all these conditions \eqref{1.5} are satisfied locally near to the  identity $F=\id$ for constitutive relations  that linearize to
\begin{align}
	\sigma_{\rm lin}(\varepsilon)=2\, \mu\, \varepsilon +\lambda\, \tr(\varepsilon)\, \id,
\end{align}
with $\varepsilon=\sym\, \D u$ the infinitesimal strain tensor and  $u(x)=\varphi(x)-x$ the displacement and $\mu>0, 2\, \mu+3\, \lambda>0$. Our interest, therefore, lies in checking the conditions in \eqref{1.5} {\bf everywhere} locally.

	Due to isotropy, the right hand side conditions in \eqref{1.5} can be suitably checked in the principal stress-principal strain representation, see  \cite{MartinVossGhibaNeff,Ogden83,hill1968constitutivea,hill1968constitutiveb,hill1970constitutive} and the  Appendix.
	
	We will show the procedure exemplarily for a constitutive choice, a toy case only to illustrate the calculations:
	\begin{itemize}
		\item the energy we take is the quadratic Biot-energy
		\begin{align}
			W_{\rm Biot} (F)&=\mu\|U-\id_3\|^2+\frac{\lambda}{2}[\tr(U-\id_3)]\notag,
		\end{align}
		where $\mu$ and $\lambda$ are the Lam\'e constants.
		\item this implies for the respective stress tensors
		\begin{align}
			T_{\rm Biot}(U)&={\rm D}_U\widetilde{W}(U)=2\,\mu(U-\id_3)+\lambda\,\tr(U-\id_3)\,\id_3,\notag\\
		\sigma(V)&=\frac{1}{\det F}[{\rm D}_F W(F)]\,F^T=\frac{1}{\det F}\, V\,\left[2\,\mu(V-\id_3)+\lambda\,\tr(V-\id_3)\,\id_3 \right],\notag\\
		\tau(V)&=\det V\cdot\,\sigma=V\,\left[2\,\mu(V-\id_3)+\lambda\,\tr(V-\id_3)\,\id_3 \right],\\
		\widehat{T}_{\rm Biot}(\log U)&=2\,\mu(e^{\log U}-\id_3)+\lambda\,\tr(e^{\log U}-\id_3)\,\id_3,\notag\\
		\widehat{\sigma}(\log V)&=\frac{1}{\det e^{\log V}}\, e^{\log V}\,\left[2\,\mu(e^{\log V}-\id_3)+\lambda\,\tr(e^{\log V}-\id_3)\,\id_3 \right],\notag\\
		\widehat{\tau}(\log V)&= e^{\log V}\,\left[2\,\mu(V-\id_3)+\lambda\,\tr(e^{\log V}-\id_3)\,\id_3 \right].
	\end{align}
		\item switching to the principal stress-principal stretch relation, this entails
		\begin{align}
			\begin{pmatrix}T_1(\lambda_1,\lambda_2,\lambda_3)\\
				T_2(\lambda_1,\lambda_2,\lambda_3)\\
				T_3(\lambda_1,\lambda_2,\lambda_3)
				\end{pmatrix}&=\left(
				\begin{array}{c}
				2 \left(\lambda _1-1\right) \mu +\lambda  \left(\lambda _1+\lambda _2+\lambda _3-3\right) \\
				2 \left(\lambda _2-1\right) \mu +\lambda  \left(\lambda _1+\lambda _2+\lambda _3-3\right) \\
				2 \left(\lambda _3-1\right) \mu +\lambda  \left(\lambda _1+\lambda _2+\lambda _3-3\right) \\
				\end{array}
				\right),\notag\\
				\begin{pmatrix}\sigma_1(\lambda_1,\lambda_2,\lambda_3)\\
					\sigma_2(\lambda_1,\lambda_2,\lambda_3)\\
					\sigma_3(\lambda_1,\lambda_2,\lambda_3)
				\end{pmatrix}&=\left(
				\begin{array}{c}
				\frac{2 \left(\lambda _1-1\right) \mu +\lambda  \left(\lambda _1+\lambda _2+\lambda _3-3\right)}{\lambda _2 \lambda _3} \\
				\frac{2 \left(\lambda _2-1\right) \mu +\lambda  \left(\lambda _1+\lambda _2+\lambda _3-3\right)}{\lambda _1 \lambda _3} \\
				\frac{2 \left(\lambda _3-1\right) \mu +\lambda  \left(\lambda _1+\lambda _2+\lambda _3-3\right)}{\lambda _1 \lambda _2} \\
				\end{array}
				\right),\notag\\
					\begin{pmatrix}\tau_1(\lambda_1,\lambda_2,\lambda_3)\\
					\tau_2(\lambda_1,\lambda_2,\lambda_3)\\
					\tau_3(\lambda_1,\lambda_2,\lambda_3)
				\end{pmatrix}&=\left(
				\begin{array}{c}
				\lambda _1 \left(2 \left(\lambda _1-1\right) \mu +\lambda  \left(\lambda _1+\lambda _2+\lambda _3-3\right)\right) \\
				\lambda _2 \left(2 \left(\lambda _2-1\right) \mu +\lambda  \left(\lambda _1+\lambda _2+\lambda _3-3\right)\right) \\
				\lambda _3 \left(2 \left(\lambda _3-1\right) \mu +\lambda  \left(\lambda _1+\lambda _2+\lambda _3-3\right)\right) \\
				\end{array}
				\right),\\
					\begin{pmatrix}\widehat{T}_1(\beta_1,\beta_2,\beta_3)\\
					\widehat{T}_2(\beta_1,\beta_2,\beta_3)\\
					\widehat{T}_3(\beta_1,\beta_2,\beta_3)
				\end{pmatrix}&=\left(
				\begin{array}{c}
				\left(e^{\beta _1}+e^{\beta _2}+e^{\beta _3}-3\right) \lambda +2 \left(e^{\beta _1}-1\right) \mu  \\
				\left(e^{\beta _1}+e^{\beta _2}+e^{\beta _3}-3\right) \lambda +2 \left(e^{\beta _2}-1\right) \mu  \\
				e^{\beta _3} (\lambda +2 \mu )+\left(e^{\beta _1}+e^{\beta _2}-3\right) \lambda -2 \mu  \\
				\end{array}
				\right),
				\notag\\
				\begin{pmatrix}\widehat{\sigma}_1(\beta_1,\beta_2,\beta_3)\notag\\
					\widehat{\sigma}_2(\beta_1,\beta_2,\beta_3)\\
					\widehat{\sigma}_3(\beta_1,\beta_2,\beta_3)
				\end{pmatrix}&=\left(
				\begin{array}{c}
				e^{-\beta _2-\beta _3} \left(\left(e^{\beta _1}+e^{\beta _2}+e^{\beta _3}-3\right) \lambda +2 \left(e^{\beta _1}-1\right) \mu \right) \\
				e^{-\beta _1-\beta _3} \left(\left(e^{\beta _1}+e^{\beta _2}+e^{\beta _3}-3\right) \lambda +2 \left(e^{\beta _2}-1\right) \mu \right) \\
				e^{-\beta _1-\beta _2} \left(e^{\beta _3} (\lambda +2 \mu )+\left(e^{\beta _1}+e^{\beta _2}-3\right) \lambda -2 \mu \right) \\
				\end{array}
				\right),
				\notag\\
				\begin{pmatrix}\widehat{\tau}_1(\beta_1,\beta_2,\beta_3)\\
					\widehat{\tau}_2(\beta_1,\beta_2,\beta_3)\\
					\widehat{\tau}_3(\beta_1,\beta_2,\beta_3)
				\end{pmatrix}&=\left(
				\begin{array}{c}
				e^{\beta _1} \left(\left(e^{\beta _1}+e^{\beta _2}+e^{\beta _3}-3\right) \lambda +2 \left(e^{\beta _1}-1\right) \mu \right) \\
				e^{\beta _2} \left(\left(e^{\beta _1}+e^{\beta _2}+e^{\beta _3}-3\right) \lambda +2 \left(e^{\beta _2}-1\right) \mu \right) \\
				e^{\beta _3} \left(e^{\beta _3} (\lambda +2 \mu )+\left(e^{\beta _1}+e^{\beta _2}-3\right) \lambda -2 \mu \right) \\
				\end{array}
				\right),\notag
		\end{align}
		where $T_i$, $\sigma_i$, $\tau_i$ denote the principal Biot stresses, the principal Cauchy stresses and the principal Kirchhoff stresses, respectively, $\beta_i=\log \lambda_i$, $\widehat{T}_i(\beta_1,\beta_2,\beta_3)=T_i(\lambda_i)$,  $\widehat{\sigma}_i(\beta_1,\beta_2,\beta_3)=\sigma_i(\lambda_i)$,  $\widehat{\tau}_i(\beta_1,\beta_2,\beta_3)=\tau_i(\lambda_i)$.
	\end{itemize}
	
	Now, e.g., $\widetilde{\sigma}:=\begin{pmatrix}
		
	\sigma_1&\sigma_2& \sigma_3
	\end{pmatrix}^T: \mathbb{R}^+\times \mathbb{R}^+\times \mathbb{R}^+\to \mathbb{R}^3$ so that the differential is the Jacobian matrix ${\rm D}_\lambda\widehat{\sigma}\in \mathbb{R}^{3\times 3}$. Then we can easily check numerically $\det {\rm D}_\lambda\widetilde{\sigma}>0$ etc., similarly for the other stress tensors.
	
	For the monotonicity investigation, from \cite{MartinVossGhibaNeff,ghiba2024biot} we infer that it is enough to check the positivity of the quadratic form \begin{align}\label{pd}\langle {\rm D}_\lambda \widetilde{\sigma} (\lambda_1,\lambda_2,\lambda_3).\xi, \xi\rangle>0 \qquad \text{for all}\qquad  \xi  \in \mathbb{R}^3\setminus\{0\}.
	\end{align}
	However, ${\rm D}_\lambda \widetilde{\sigma}$ may not be symmetric (no major  symmetry of  ${\rm D}_V\sigma(V)$). Since,   due to the property of the  quadratic forms,  \eqref{pd} is equivalent to
	\begin{align}\label{pds}\langle \sym\, {\rm D}_\lambda \widetilde{\sigma} (\lambda_1,\lambda_2,\lambda_3).\xi, \xi\rangle>0 \qquad \text{for all}\qquad  \xi  \in \mathbb{R}^3\setminus\{0\},
	\end{align}
	we can we always check the positive definiteness of $\sym\, {\rm D}_\lambda \widehat{\sigma}$ via  the standard methods (e.g., Sylvester's criterion). In order to establish \eqref{pds}, note that due to the Bendixson Theorem \cite{bendixson1902racines}
	\begin{align}
		\sym\, {\rm D}_\lambda \widetilde{\sigma}\in {\rm Sym}^{++}(3)\qquad  \Longrightarrow\qquad \det {\rm D}_\lambda \widetilde{\sigma}>0.
	\end{align}

	\restylefloat*{figure}
	
	\begin{figure}[!h]
		\centering
		\begin{minipage}{.25\textwidth}
			\includegraphics[scale=0.25]{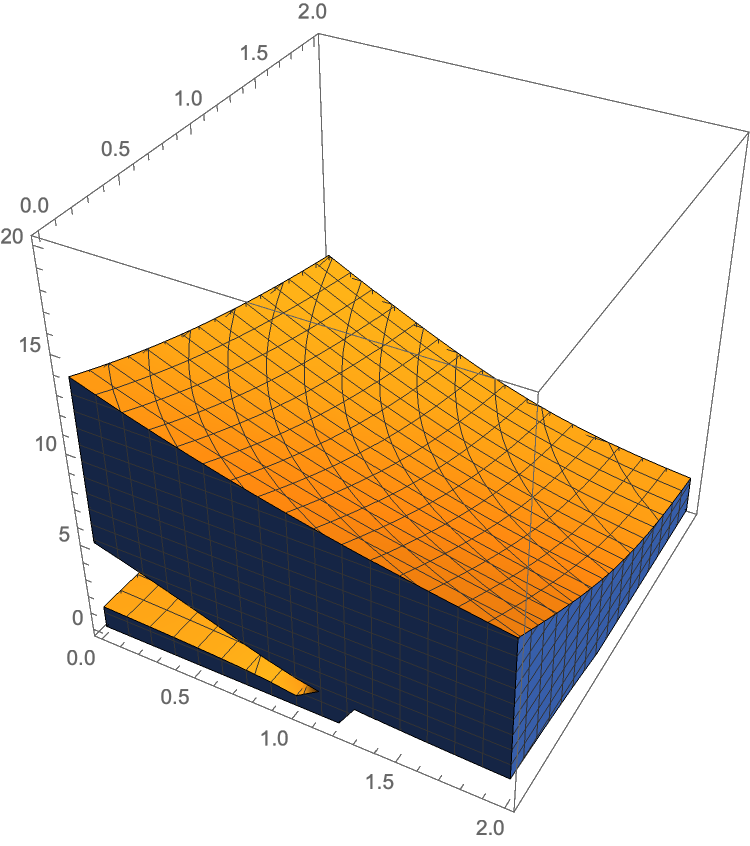}
			\caption{\footnotesize{$V\mapsto \sigma(V)$ local invertibility for the Biot energy. Local invertibility is not satisfied. }}
			\label{1}
		\end{minipage}
		\qquad \quad
		\begin{minipage}{.25\textwidth}
			\includegraphics[scale=0.25]{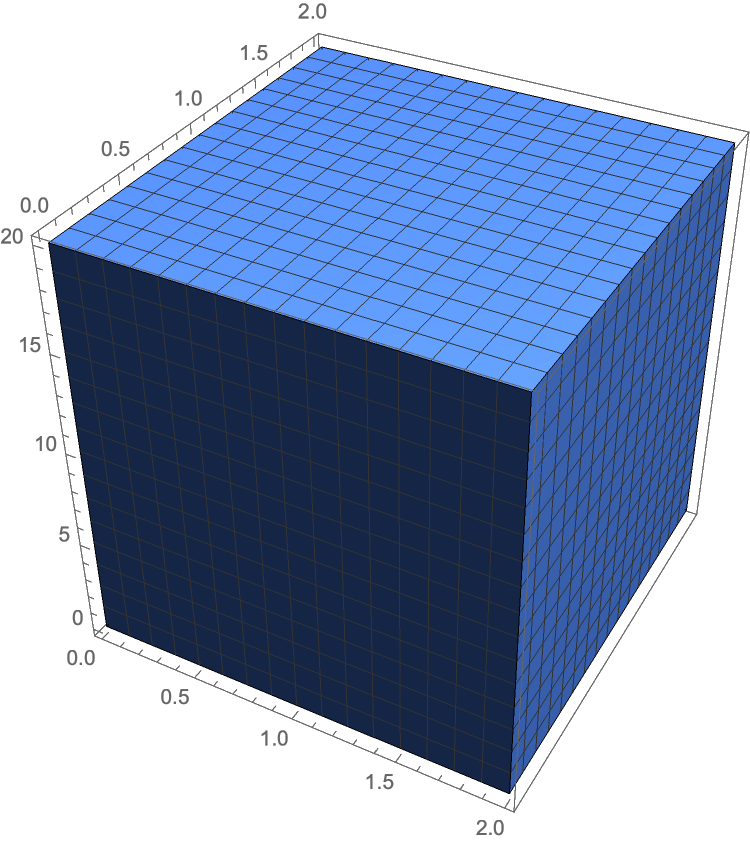}
			\caption{\footnotesize{$U\mapsto T_{\rm Biot}(U)$ local invertibility  for the Biot energy. Local invertibility is  satisfied.}}
			\label{2}
		\end{minipage}
		\qquad \quad
		\begin{minipage}{.25\textwidth}
			\includegraphics[scale=0.25]{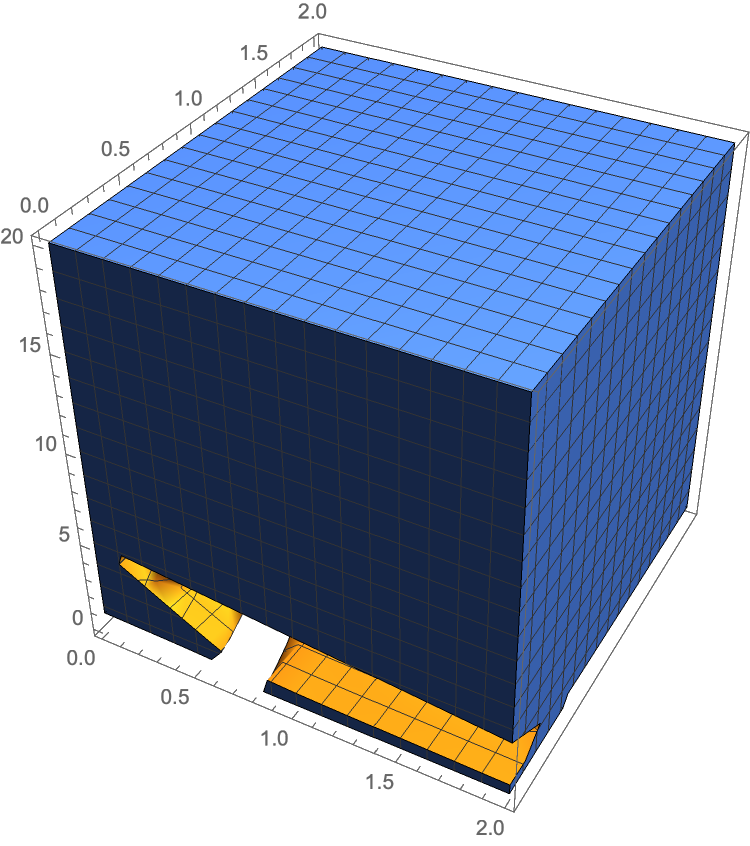}
			\caption{\footnotesize{$V\mapsto \tau(V)$ local invertibility for the Biot energy. Local invertibility is not satisfied.}}
			\label{3}
		\end{minipage}\\
		\centering
		\begin{minipage}{.25\textwidth}
			\includegraphics[scale=0.25]{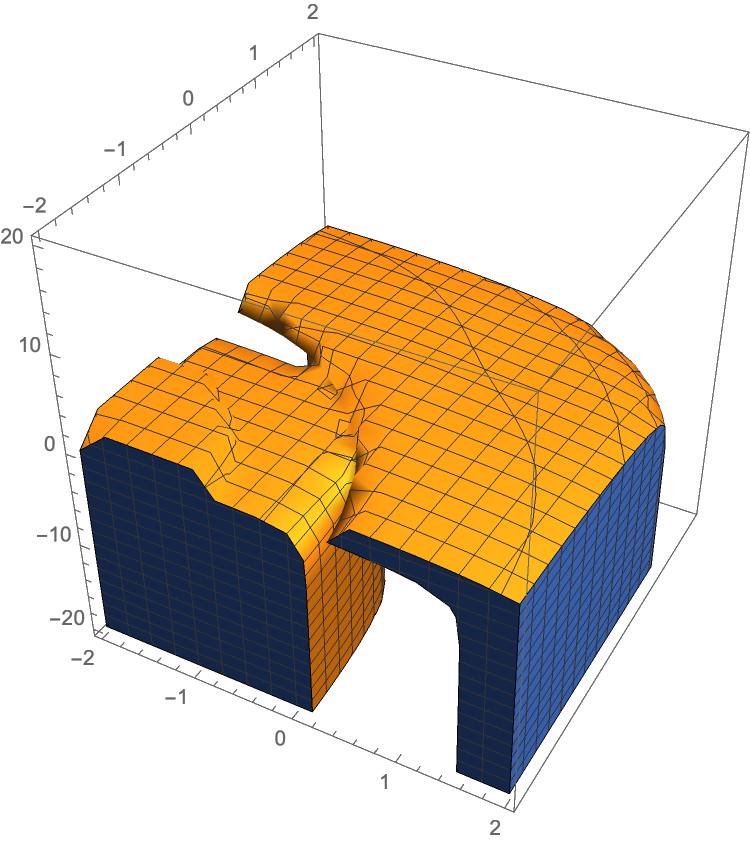}
			\caption{\footnotesize{$\log V\mapsto \widehat{\sigma}(\log V)$ local invertibility  for the Biot energy. Local invertibility is not satisfied.}}
			\label{4}
		\end{minipage}
		\qquad \quad
		\begin{minipage}{.25\textwidth}
			\includegraphics[scale=0.25]{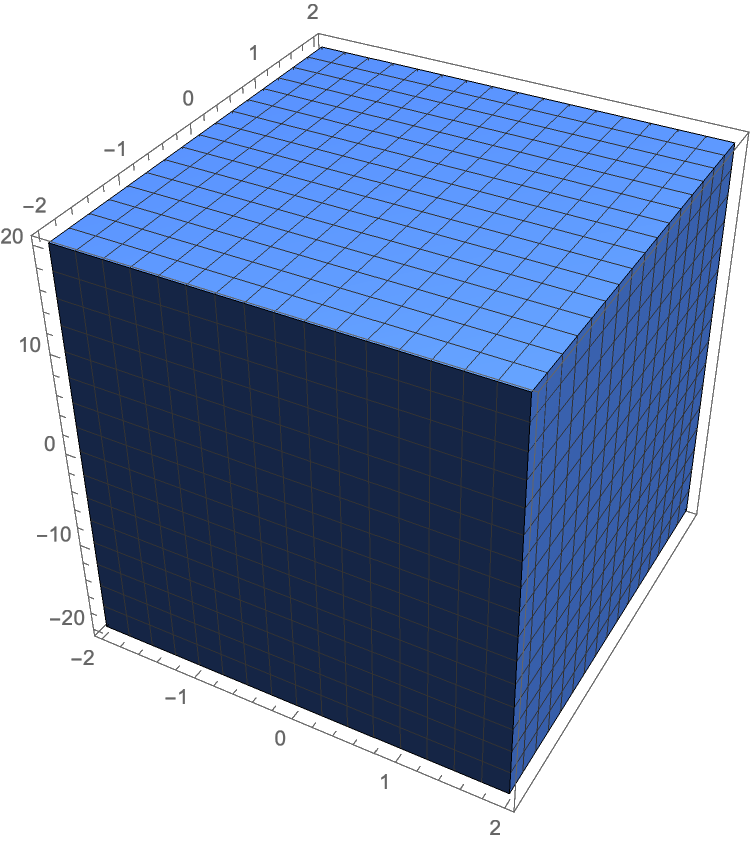}
			\caption{\footnotesize{$\log U\mapsto \widehat{T}_{\rm Biot}(\log U)$ local invertibility  for the Biot energy. Local invertibility is satisfied.}}
			\label{5}
		\end{minipage}
		\qquad \quad
		\begin{minipage}{.25\textwidth}
			\includegraphics[scale=0.25]{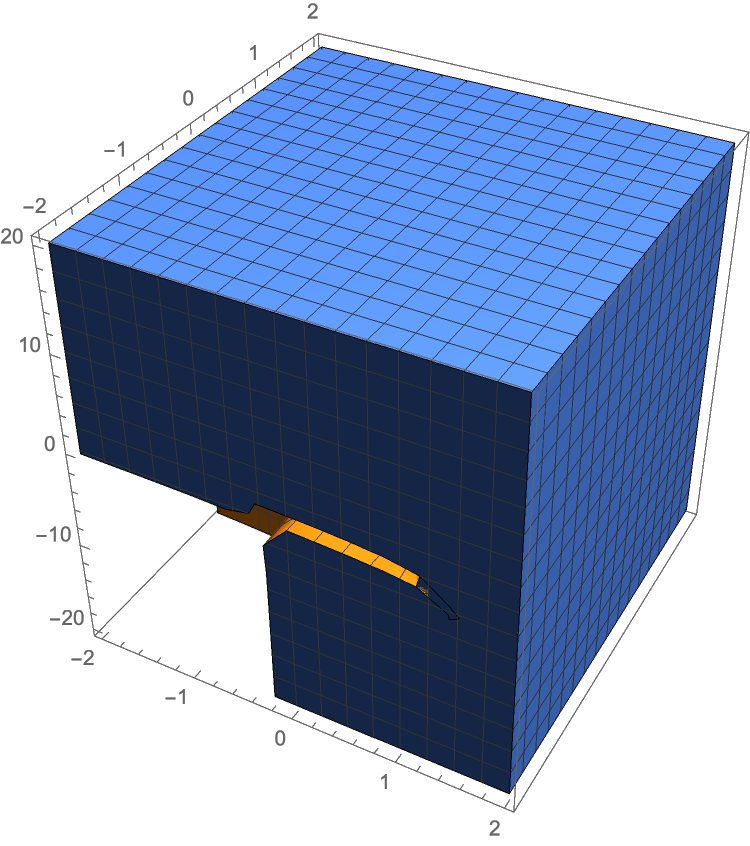}
			\caption{\footnotesize{$\log V\mapsto \widehat{\tau}(\log V)$ local invertibility  for the Biot energy. Local invertibility is not satisfied.}}
			\label{6}
		\end{minipage}
		\\
		\centering
		\begin{minipage}{.25\textwidth}
			\includegraphics[scale=0.25]{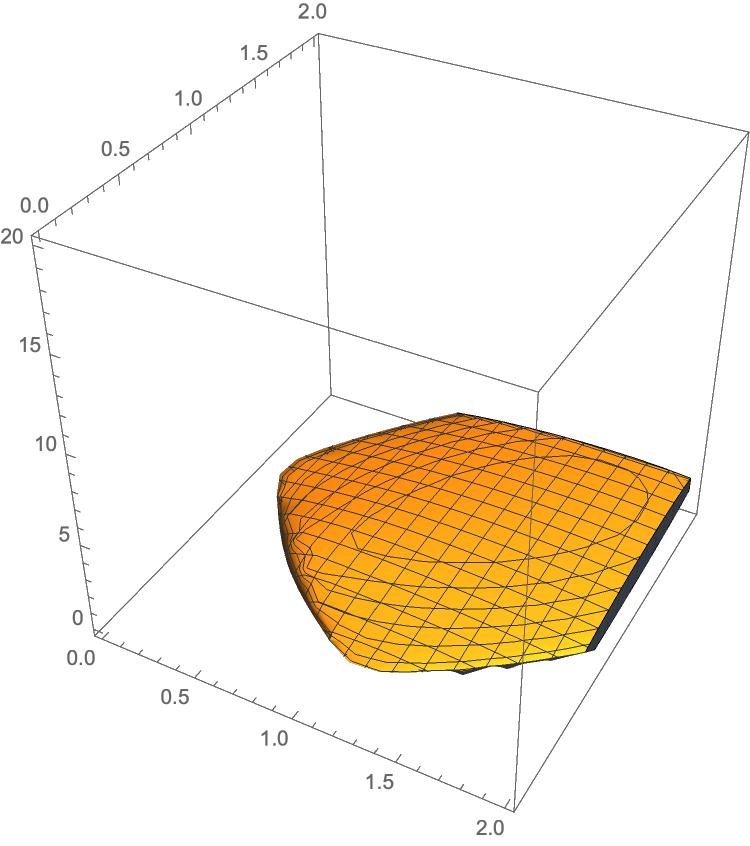}
			\caption{\footnotesize{$V\mapsto \sigma(V)$ local monotonicity  for the Biot energy. Local monotonicity is not satisfied.}}
			\label{7}
		\end{minipage}
		\qquad \quad
		\begin{minipage}{.25\textwidth}
			\includegraphics[scale=0.25]{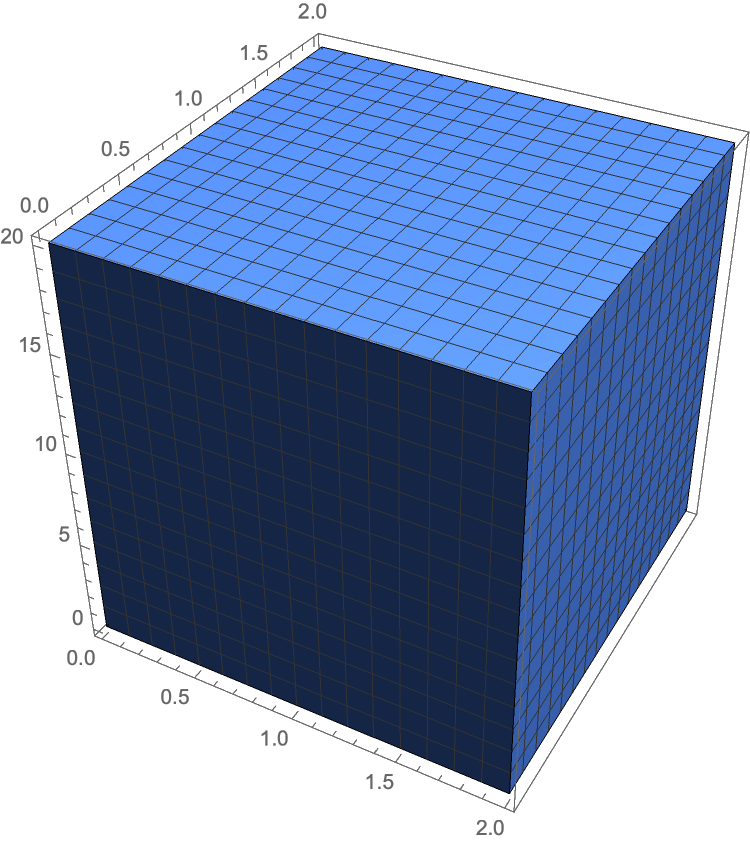}
			\caption{\footnotesize{$U\mapsto T_{\rm Biot}(U)$ local monotonicity  for the Biot energy. Local monotonicity is  satisfied.}}
			\label{8}
		\end{minipage}
		\qquad \quad
		\begin{minipage}{.25\textwidth}
			\includegraphics[scale=0.25]{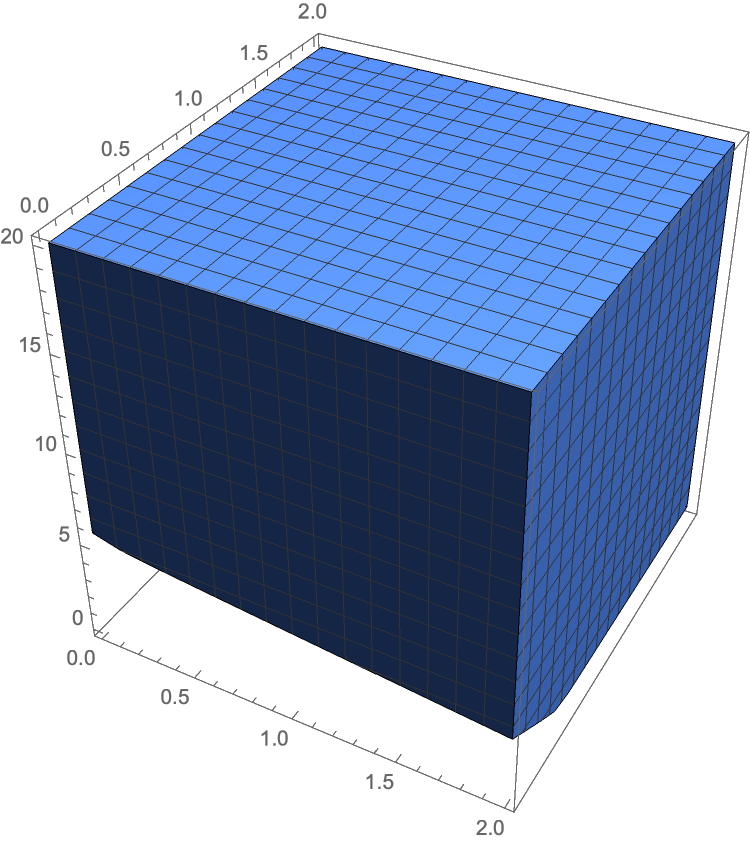}
			\caption{\footnotesize{$V\mapsto \tau(V)$ local monotonicity  for the Biot energy. Local monotonicity is  not satisfied.}}
			\label{9}
		\end{minipage}\\
		\centering
		\begin{minipage}{.25\textwidth}
			\includegraphics[scale=0.25]{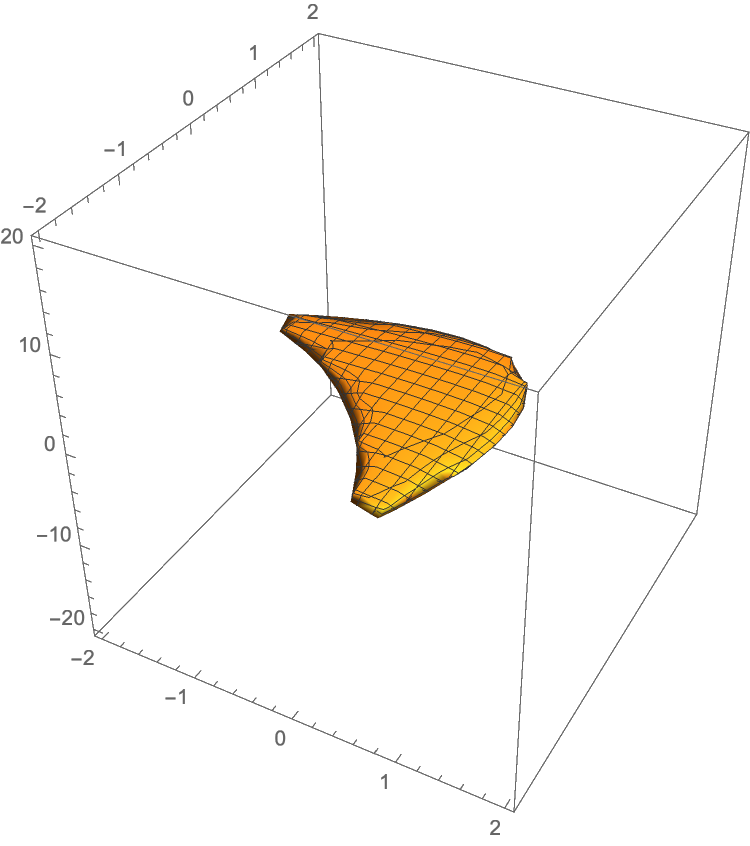}
			\caption{\footnotesize{$\log V\mapsto \widehat{\sigma}(\log V)$ local monotonicity  for the Biot energy. Local monotonicity is not satisfied.}}
			\label{10}
		\end{minipage}
		\qquad \quad
		\begin{minipage}{.25\textwidth}
			\includegraphics[scale=0.25]{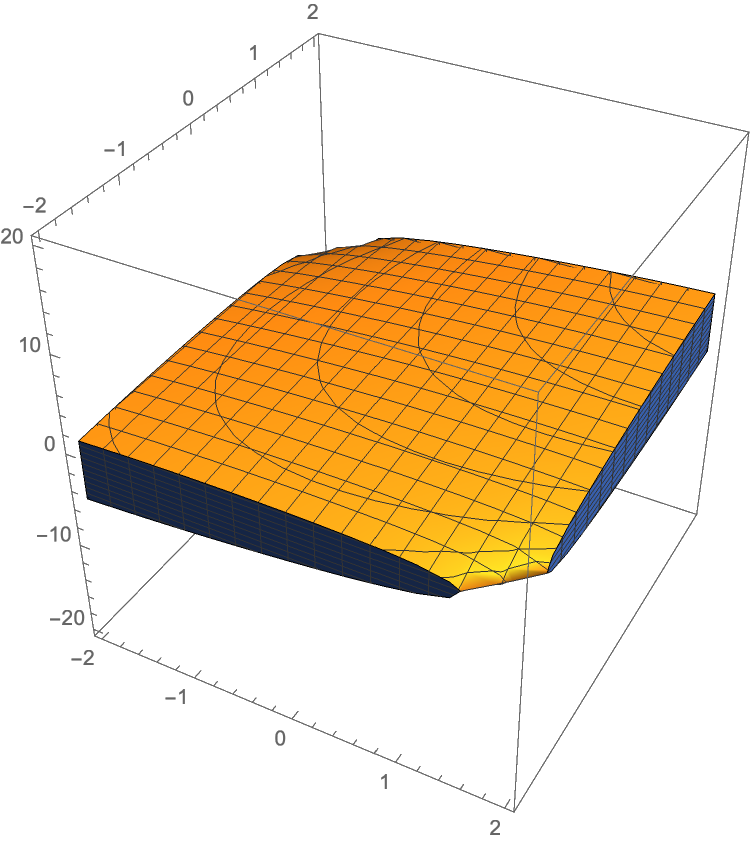}
			\caption{\footnotesize{$\log U\mapsto \widehat{T}_{\rm Biot}(\log U)$ local monotonicity  for the Biot energy. Local monotonicity is not satisfied.}}
			\label{11}
		\end{minipage}
		\qquad \quad
		\begin{minipage}{.25\textwidth}
			\includegraphics[scale=0.25]{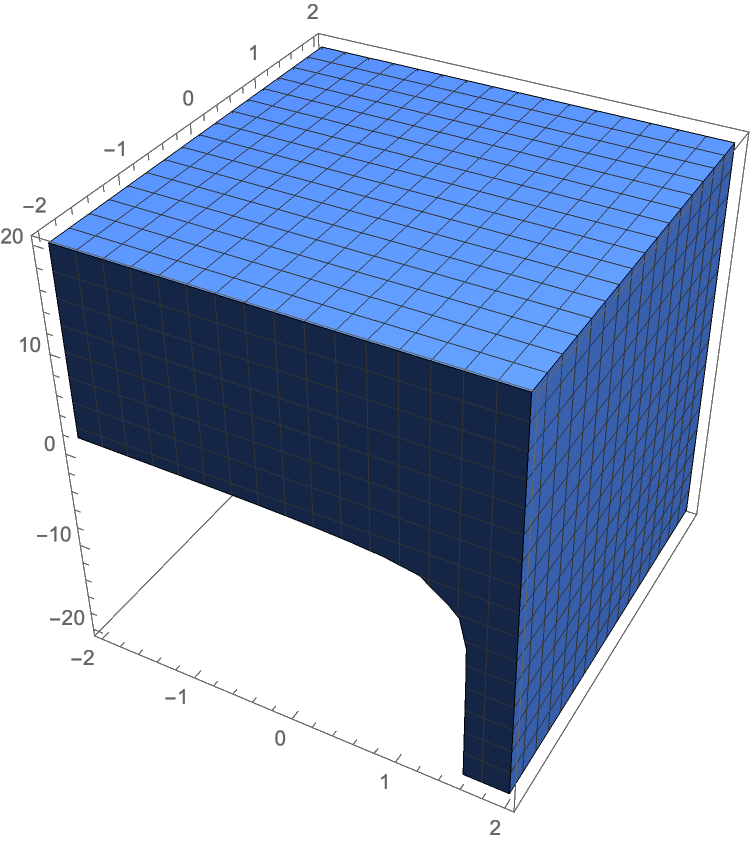}
			\caption{\footnotesize{$\log V\mapsto \widehat{\tau}(\log V)$ local monotonicity  for the Biot energy. Local monotonicity is not satisfied.}}
			\label{12}
		\end{minipage}
	\end{figure}
	
	In the following we explicitly disregard the pair $C\mapsto S_2(C)$ in terms of invertibility and monotonicity properties as these do not contribute to our understanding of constitutive properties, notwithstanding some useful results for energies that are convex in $C$, see \cite{gao2015convexity}.

	\section{An incomplete and biased list of isotropic elastic energies for compressible materials}
For different types of compressible materials or for various behaviour we want to capture in the modelling process, one must  choose an appropriate elastic energy. A set of possible isotropic energies proposed for compressible elastic materials in the literature is (this choice is biased):
	\subsection{The Biot energy}\label{Biot-energy}
		\begin{align}
		W_{\rm{Biot}}(F)\,&=\, \mu\,\lVert U-\id_3\rVert^2+\frac{\lambda}{2}\,[\tr(U-\id_3)]^2
		\,=\,\, \mu\,\lVert V-\id_3\rVert^2+\frac{\lambda}{2}\,[\tr(V-\id_3)]^2.
		\end{align}
		
			\subsection{The   slightly compressible  generalised Biot material incorporating a volumetric-isochoric split which does not explode}\label{Biotsde}
		\begin{align}
			W_{\rm Biot}^{\rm vol+iso }(F):&=\frac{\mu}{2}\| \frac{U}{\det U^{1/3}}-\id\|^2+\frac{\kappa}{2} (\det U-1)^2,
			\notag\end{align}
		which additively separates the isochoric and volumetric contributions.
		\subsection{Another  slightly compressible  generalised Biot material incorporating a volumetric-isochoric split which  explodes}\label{Biotse}
		\begin{align}
			W_{\rm Biot}^{\rm vol+iso }(F):&=\frac{\mu}{2}\| \frac{U}{\det U^{1/3}}-\id\|^2+\frac{\kappa}{2\, \widehat{k}}\,\left(e^{\widehat{k}(\log \det F)^2}-1\right).
		\end{align}

		\subsection{The  Saint-Venant--Kirchhoff energy}\label{SV-energy}
		\begin{align}W_{\rm SVK}(F)&={\mu}\, \lVert E\rVert^2+\frac{\lambda}{2}[\tr(E)]^2=\frac{\mu}{4}\, \lVert U^2-\id_3\rVert^2+\frac{\lambda}{8}[\tr(U^2-\id_3)]^2=\frac{\mu}{4}\, \lVert C-\id_3\rVert^2+\frac{\lambda}{8}[\tr(C-\id_3)]^2\notag\\
			&=\frac{1}{4}\,\left[\frac{1}{2}(\lambda+2\mu)I_1^2(E)-2\, \mu\, I_2(E)\right],\end{align}
		where $E=\frac{1}{2}(C-\id_3), \ C=F^TF$.

		\subsection{The slightly  compressible  Mooney-Rivlin energy incorporating a volumetric-isochoric split which explodes}\label{MRsc-energy}
		\begin{align}\label{sMR}
	W_{\rm MR}^{\rm vol+iso}(F)&=\frac{a }{2}\left(\left\|\frac{F}{\det F^{1/3}}\right\|^2-3\right)+\frac{b }{2}\left(\left\|\Cof(\frac{F}{\det F^{1/3}})\right\|^2-3\right)+\frac{\kappa}{2\, \widehat{k}}\,\left(e^{\widehat{k}(\log \det F)^2}-1\right)\\
	&=\frac{a }{2}\left(\left\|\frac{V}{\det V^{1/3}}\right\|^2-3\right)+\frac{b }{2}\left(\left\|\Cof(\frac{V}{\det V^{1/3}})\right\|^2-3\right)+\frac{\kappa}{2\, \widehat{k}}\,\left(e^{\widehat{k}(\log \det F)^2}-1\right), \quad   \kappa>0,\notag
		\end{align}
		where
		$\mu=a +b $ is the shear modulus and $\kappa=\frac{3\lambda+2\mu}{3}$ is the bulk modulus.
		
			\subsection{The slightly  compressible  Mooney-Rivlin energy incorporating a volumetric-isochoric split which does not explode}\label{MRsc-energy0}
		\begin{align}\label{sMR}
			W_{\rm MR}^{\rm vol+iso}(F)&=\frac{a }{2}\left(\left\|\frac{F}{\det F^{1/3}}\right\|^2-3\right)+\frac{b }{2}\left(\left\|\Cof(\frac{F}{\det F^{1/3}})\right\|^2-3\right)+\frac{\kappa}{2}\,\left( \det F-1\right)^2\\
			&=\frac{a }{2}\left(\left\|\frac{V}{\det V^{1/3}}\right\|^2-3\right)+\frac{b }{2}\left(\left\|\Cof(\frac{V}{\det V^{1/3}})\right\|^2-3\right)+\frac{\kappa}{2}\,\left( \det F-1\right)^2, \quad    \kappa>0,\notag
		\end{align}
		where
		$\mu=a +b $ is the shear modulus and $\kappa=\frac{3\lambda+2\mu}{3}$ is the bulk modulus.
	
			\subsection{The slightly compressible exponentiated Mooney energy }\label{expslightlyMooney-energy}
		
		\begin{align}\label{sMR}
			W_{\rm MR}^{\rm vol+iso}(F)&=\frac{\mu}{e^{3/2}}e^{\left(\frac{a }{2}\left\|\frac{F}{\det F^{1/3}}\right\|^2+\frac{1-a }{2}\left\|\Cof(\frac{F}{\det F^{1/3}})\right\|^2\right)}+\frac{\kappa}{2\, \widehat{k}}\,e^{\widehat{k}(\log \det F)^2}-\frac{2\,\mu\,+ \kappa }{2 \widehat{k}}\\
			&=\frac{\mu}{e^{3/2}}e^{\left(\frac{a }{2}\left\|\frac{V}{\det V^{1/3}}\right\|^2+\frac{1-a }{2}\left\|\Cof(\frac{V}{\det V^{1/3}})\right\|^2\right)}+\frac{\kappa}{2\, \widehat{k}}\,e^{\widehat{k}(\log \det F)^2}-\frac{2\,\mu\, +\kappa }{2 \widehat{k}},\qquad    \  \widehat{k}>0,\  c>0.\notag
		\end{align}
			
	\subsection{The 2-parameter Mooney-Rivlin material model \cite{eisentrager2025implementation}}\label{2parMR}
	The two-parameter model of Moooney \cite{mooney1940theory} and Rivlin \cite{rivlin1948large} is
	\begin{align}
		W(F)=\frac{C_{10}}{2}(\|F\|^2-3)+\frac{C_{01}}{2}(\|\Cof F\|^2-3)+\frac{\lambda}{4}((\det F)^2-1-2\log \det F)-(C_{10}+2\,C_{01})\log \det F.
	\end{align}
	
	\subsection{The   slightly compressible  generalised Neo-Hooke material incorporating a volumetric-isochoric split which does not explode}\label{NHsc-energy}
		\begin{align}
		W_{\rm NH}^{\rm iso}(F)&=\frac{\mu}{2}\langle \frac{B}{\det B^{1/3}}-\id,\id\rangle+\frac{\kappa}{2} (\det F-1)^2=\frac{\mu}{2}\langle \frac{V^2}{\det V^{2/3}}-\id,\id\rangle+\frac{\kappa}{2}\, (\det F-1)^2\\
		&=\frac{\mu}{2}\left( \frac{\|F\|^2}{\det F^{2/3}}-3\right)+\frac{\kappa}{2}\, (\det F-1)^2,
		\notag\end{align}
		 which additively separates the isochoric and volumetric contributions.
		\subsection{Another  slightly compressible  generalised Neo-Hooke material incorporating a volumetric-isochoric split which  explodes}\label{NHvi-energy}
		\begin{align}
			W_{\rm NH}^{\rm iso}(F)
&=\frac{\mu}{2}\left( \frac{\|F\|^2}{\det F^{2/3}}-3\right)+\frac{\kappa}{2\, \widehat{k}}\,\left(e^{\widehat{k}(\log \det F)^2}-1\right).
		\end{align}

		\subsection{The original Ciarlet-Geymonat energy for compressible materials \cite{glugegraphical} by}\label{CG-energy}
		\begin{align}
		W_{\rm CG}(F)&=\frac{\mu}{2}\left[\|F\|^2-2\,\log (\det F)-3\right]+\frac{\lambda}{4}\left[\,(\det F)^2-2\,\log (\det F)-1\right]\\
		&=\frac{\mu}{2}\left[\tr (V^2)-2\,\log (\det V)-3\right]+\frac{\lambda}{4}\left[\,(\det V)^2-2\,\log (\det V)-1\right].\notag
		\end{align}
		
			\subsection{The simplified  Ciarlet-Geymonat energy for compressible materials}\label{CG-energys}
		\begin{align}
			W_{\rm CG}(F)&=\frac{\mu}{2}\left[\|F\|^2-2\,\log (\det F)-3\right]=\frac{\mu}{2}\left[\tr (V^2)-2\,\log (\det V)-3\right].\notag
		\end{align}
This  energy  is the one parameter energy convex in $U$.

		\subsection{The original Simo-Pister energy for compressible materials \cite{simo1984remarks} }\label{SP}
		\begin{align}
			W_{\rm SP}(F)&=\frac{\mu}{2}\left[\|F\|^2-2\,\log (\det F)-3\right]+\frac{\lambda}{2}\,\log^2 (\det F).\notag
		\end{align}
		
		\subsection{The modified Simo-Pister energy for compressible materials}\label{mSP}
		\begin{align}
			W_{\rm SP}(F)&=\frac{\mu}{2}\left[\|F\|^2-2\,\log (\det F)-3\right]+\frac{\lambda}{2}\,\left(e^{\log^2 (\det F)}-1\right).\notag
		\end{align}
		\subsection{The slightly compressible Ogden-energy   obtained  using the volumetric-isochoric split}\label{Ogdenvi-energy}
		\begin{align}
		W_{\rm Ogden}^{\rm slightly}(F)=\sum_{k=1}^N\frac{\mu_k}{\alpha_k}\left(\overline{\lambda}_1^{\alpha_k}
		+\overline{\lambda}_2^{\alpha_k}+\overline{\lambda}_3^{\alpha_k}-3\right)+\frac{\kappa}{2\, \widehat{k}}\,\left(e^{\widehat{k}(\log \det F)^2}-1\right),  \qquad \quad  \frac{\mu_k}{\alpha_k}>0,
		\end{align}
	where $	
	\overline{\lambda}_i=\frac{\lambda_i}{(\det F)^{1/3}}$	denote the modified stretch ratios and $\mu=\frac{1}{2}\sum_{k=1}^N \alpha_k\cdot \mu_k$ is the initial shear modulus.
		
		The Ogden model relies on an iterative nonlinear curve fitting method while $\kappa$ is the bulk modulus. For this reason and also due to the several coefficients involved, the adequate fitting of the Ogden model is highly dependent on the initial values proposed before the curve fittings begins.
		
		A special case of the slightly compressible Ogden energy is the slightly  compressible Mooney-Rivlin model with  volumetric-isochoric split, see \eqref{sMR}.

		\subsection{The quadratic Hencky model (volumetric-isochoric split) \cite{Neff_Osterbrink_Martin_hencky13,neff2020axiomatic,graban2019commented,neff2016rediscovering}}\label{Hencky-energy}
		\begin{align}
		W_{\rm H}(F)&=\mu\|\dev\log U\|^2+\frac{\kappa}{2}[\tr(\log U)]^2=\mu\|\dev\log V\|^2+\frac{\kappa}{2}[\tr(\log V)]^2\notag\\&=\mu\|\log \frac{U}{\det U^{1/3}} \|^2+\frac{\kappa}{2}[\tr(\log U)]^2 =\mu\|\!\log V\|^2+\frac{\lambda}{2}[\tr(\log V)]^2.
		\end{align}

			\subsection{The  Fung-type energy for slightly compressible materials (volumetric-isochoric split)}\label{Fuvi-energy}
				\begin{align}
			W(F)=\frac{\mu}{ 2k}\left[e^{k\,(\|\frac{V}{(\det V)^{1/3}}\|^2-3)}-1\right]+\frac{\kappa}{2\, \widehat{k}}\,\left(e^{\widehat{k}(\log \det F)^2}-1\right), \qquad k>0.
			\end{align}

				This energy  is often \cite{fung1979inversion,beatty1987topics} used in the biomechanics
			literature to describe the nonlinear strain stiffening elastic response of biological tissues
		\subsection{The hyper Hooke energy  (exponentiation of the Hencky-energy)}\label{expHencky-energy}
		\begin{align}
		W_{\rm hyper}(F)&=\frac{\mu}{k}\left(e^{k\,\|\dev\log U\|^2+\frac{\widehat{k}}{2}[\tr(\log U)]^2}-1\right)=\frac{\mu}{k}\left(e^{k\,\|\dev\log V\|^2+\frac{\widehat{k}}{2}[\tr(\log V)]^2}-1\right),
		\end{align}
where  $\lambda=\frac{\mu \left(\widehat{k}-\frac{2 k}{3}\right)}{k}$.
		\subsection{The hyper Hooke isochoric  energy term}\label{hyperHookei-energy}
	\begin{align}
	W_{\rm H}(F)&=\frac{\mu}{k}\left(e^{k\|\dev\log U\|^2}-1\right)=\frac{\mu}{k}\left(e^{k\|\dev\log V\|^2}-1\right),
	\end{align}
	i.e., $\kappa=0$.
	
	\subsection{The simplified hyper Hooke energy}\label{shyperHooke-energy}
	\begin{align}
	W_{\rm H}(F)&=\frac{\mu}{k}\left(e^{k\,\|\!\log U\|^2}-1\right)=\frac{\mu}{k}\left(e^{k\,\|\!\log V\|^2}-1\right)=\frac{\mu}{k}\left(e^{k\,\|\dev\log V\|^2+\frac{k}{3}\,[\tr (\log V)]^2 }-1\right),
	\end{align}
	i.e., $\lambda=0$.
	\subsection{The volumetric-isochoric split exponentiated Hencky-energy}\label{expHenckyvi-energy}
		\begin{align}\label{the}
	W_{_{\rm eH}}(F)&=
	\underbrace{\frac{\mu}{k}\,\left(e^{k\,\|\dev\log\,U\|^2}-1\right)+\frac{\kappa}{2\widehat{k}}\,\left(
		e^{\widehat{k}\,(\tr(\log U){)}^2}-1\right)}_{\text{volumetric-isochoric split}}=	 \frac{\mu}{k}\,\left(e^{k\,\|\dev\log\,V\|^2}-1\right)+\frac{\kappa}{2\widehat{k}}\,\left(
		e^{\widehat{k}\,(\tr(\log V){)}^2}-1\right).
	\end{align}
	
	\subsection{The non volumetric-isochoric split exponentiated Hencky-energy}\label{expHenckynvi-energy}
	\begin{align}\label{the}
		W_{_{\rm eHH}}(F)&=
	\frac{\mu}{k}\,\left(e^{k\,\|\!\log\,U\|^2}-1\right)+\frac{\lambda}{2\widehat{k}}\,
			\left(e^{\widehat{k}\,(\tr(\log U){)}^2}-1\right)=	 \frac{\mu}{k}\,\left(e^{k\,\|\!\log\,V\|^2}-1\right)+\frac{\lambda}{2\widehat{k}}\,\left(
		e^{\widehat{k}\,(\tr(\log V){)}^2}-1\right).
	\end{align}

	\subsection{The Bazant energy } \label{Bazant-energy}
	\begin{align}
W_{\rm Bazant}(F)=	\frac{\mu}{4}\,\|V-V^{-1}\|^2+\frac{\lambda}{8}\,[\tr (V-V^{-1})]^2.
	\end{align}
	We remark the tension-compression symmetry of this energy, i.e., invariance under $V\mapsto V^{-1}$.
		\subsection{The slightly compressible Bazant energy  (volumetric-isochoric split) and tension-compression symmetry}\label{Bazantsc-energy}
	\begin{align}
	W_{\rm Bazant}(F)=	\frac{\mu}{4}\,\|\frac{V}{\det V^{1/3}}-\left(\frac{V}{\det V^{1/3}}\right)^{-1}\|^2+\frac{\lambda}{8}(\det V-\frac{1}{\det V})^2.
	\end{align}
\subsection{The exponentiated Bazant energy}\label{expBazant-energy}
\begin{align}
W_{\rm eBazant}=\frac{\mu}{4k}\left(e^{k\,\|V-V^{-1}\|^2}-1\right)+\frac{\lambda}{8\,\widehat{k}}\left(e^{\widehat{k}\,(\det V-\frac{1}{\det V})^2}-1\right).
\end{align}

	\subsection{The  exponentiated slightly compressible Bazant energy which  does not explode}  \label{expBazantsc-energyede}
\begin{align}
	W_{\rm Bazant}(F)=\frac{\mu}{k}\left(e^{k\,\|\frac{V}{\det V^{1/3}}-\left(\frac{V}{\det V^{1/3}}\right)^{-1}\|^2}-1\right)+\frac{\kappa}{2}\left(\det F-1\right)^2.
\end{align}

	\subsection{The  exponentiated slightly compressible Bazant energy which  explodes}  \label{expBazantsc-energye}
\begin{align}
W_{\rm Bazant}(F)=\frac{\mu}{k}\left(e^{k\,\|\frac{V}{\det V^{1/3}}-\left(\frac{V}{\det V^{1/3}}\right)^{-1}\|^2}-1\right)+\frac{\kappa}{2\,\widehat{k}}\left(e^{\widehat{k}\,(\det V-\frac{1}{\det V})^2}-1\right).
\end{align}

\subsection{The Saravan energy \cite{saravanan2011adequacy} }\label{Saravan-energy}
\begin{align}
W_{\rm Saravan}(F)=\mu\, \left[\frac{\|F\|^2-3}{2}+(1-\underbrace{\det F}_{\rm null-Lagrangian})\right]=\mu\, \left[\frac{\|V\|^2-3}{2}+(1-\det V)\right].
\end{align}
This one-parameter energy is not compatible with stability in linear elasticity, but it is Legendre-Hadamard-elliptic and indeed polyconvex.

\subsection{The two parameter Signorini  model \cite{lurie2012non}}\label{Signorini}
\begin{align}
	W_{\rm Signorini}(F)&=\frac{1}{2}\sqrt{\det B}\, \left[9\lambda+5\mu-2(3\lambda+\mu)\frac{I_2(B)}{\det B}+(\mu+\lambda)\left(\frac{I_2(B)}{\det B}\right)^2\right]-\mu\\&=\frac{1}{2}\det V\, \left[9\lambda+5\mu-2(3\lambda+4\mu)\frac{\|\Cof V\|^2}{(\det V)^2}+(\mu+\lambda)\left(\frac{\|\Cof V\|^2}{(\det V)^2}\right)^2\right]-\mu.\notag
\end{align}

		\subsection{The (strain limiting)  Gent energy \cite{gent1996new} for slightly compressible materials  and volumetric-isochoric split}\label{Gentsc-energy2}
		\begin{align}
			W_{\rm Gent}^{\rm slightly}(F)&=\begin{cases}-\frac{\mu }{2}j_m\log\left(1-\frac{\|\frac{F}{(\det F)^{1/3}}\|^2-3}{j_m}\right)+\frac{\kappa}{2}\left(\frac{(\det F)^2-1}{2}-\log  (\det F)\right)& \text{if}  \qquad \|\frac{F}{(\det F)^{1/3}}\|^2<j_m+3\\
					\infty& \text{else}
				\end{cases},
		\end{align}
		where
		$j_m={i_{\rm max}-3}$ and $i_{\rm max}\gg 1$ is a given arbitrary  scalar and $\mu$ and $\kappa$ are the shear and bulk modulus, respectively.
	\subsection{Another Gent-type energy \cite{Wollner}-Wollner's energy}\label{Gte}
		\begin{align}
		W(F)&=\begin{cases}-\log(\beta-\|F\|^\beta)-\gamma\, \log \det F+\left(\gamma-\alpha\frac{3^{\alpha/2-1}}{\beta-3^\alpha/2}\right) \det F+\text{const.}&\text{if}\qquad \|F\|^\alpha<\beta\\
		\infty& \text{else}
		\end{cases},
	\end{align}
	where $\alpha\geq 1$, $\beta>3^{\alpha/2}$ and $\gamma\geq 1/4$. Asking to lead to a valid elastic law in the infinitesimal theory we have that
	\begin{align}
		\mu=\alpha\,\frac{3^{\alpha/2-1}}{\beta-3^\alpha/2}>0,\qquad\quad  \kappa=\gamma+\frac{\alpha\, 3^{\alpha/2}(\alpha\,\beta-3(\beta-3^{\alpha/2}))}{9(\beta-3^{\alpha/2})}>0.
	\end{align}This energy is easily seen to be polyconvex and satisfies TSTS-M$^{++}$ within its restricted domain of definition.
	\subsection{The exponentiated Hencky-energy with limited  compressibility}\label{expHencky-tan}
	\begin{align}\label{thetan}
		W_{_{\rm tanH}}(F)&=\begin{cases}
		\frac{\mu}{k}\,\left(e^{k\,\|\!\log\,U\|^2}-1\right)+\frac{\lambda}{2}\,
		\tan(\log \det F)^2& \text{if} \quad (\log \det F)^2<\frac{\pi}{2}\\
		\infty &\text{else}
		\end{cases}.
	\end{align}
	
	\subsection{The (strain-limiting) Benam energy for slightly compressible materials which does not explode \cite{anssari2023large}}\label{Benam-energy}
	\begin{align}
		W_{\rm Benam}=\frac{3(n-1)}{2\, n} \mu\,N\,\left[\frac{1}{3N(n-1)}(\overline{\lambda}_1^\alpha+\overline{\lambda}_2^\alpha+\overline{\lambda}_3^\alpha-3-\log \left( \frac{\overline{\lambda}_1^\alpha+\overline{\lambda}_2^\alpha+\overline{\lambda}_3^\alpha-3N}{3-3N}\right)\right]+\frac{\kappa}{2}(\det F-1)^2.
	\end{align}
	
	For compatibility with  linear elasticity one must set $\frac{4 (n-1) N}{\alpha ^2 (N n-1)}=1$.
	
	\subsection{The (strain-limiting) Benam energy for slightly compressible materials which  explodes}\label{Benam-energye}
	\begin{align}
		W_{\rm Benam}=\frac{3(n-1)}{2\, n} \mu\,N\,\left[\frac{1}{3N(n-1)}(\overline{\lambda}_1^\alpha+\overline{\lambda}_2^\alpha+\overline{\lambda}_3^\alpha-3-\log \left( \frac{\overline{\lambda}_1^\alpha+\overline{\lambda}_2^\alpha+\overline{\lambda}_3^\alpha-3N}{3-3N}\right)\right]+\frac{\kappa}{2}\left(e^{(\log \det F)^2}-1\right).
	\end{align}
	
	\subsection{The tension-compression symmetric energy  (also polyconvex)}\label{tc-energy}
	\begin{align}
	W(U)&=\frac{\mu}{4} \,\left\|\frac{U}{(\det U)^{1/2}}-\left(\frac{U}{(\det U)^{1/2}}\right)^{-1}\right\|^2+\frac{\mu+2\lambda}{4\, \widehat{k}}\,\left(e^{\widehat{k}(\log \det F)^2}-1\right).
	\end{align}

	 Note that $\frac{U}{(\det U)^{1/2}}-\left(\frac{U}{(\det U)^{1/2}}\right)^{-1}$ is a strain measure in the sense that $\frac{U}{(\det U)^{1/2}}-\left(\frac{U}{(\det U)^{1/2}}\right)^{-1}=0\Longleftrightarrow U=\id\Leftrightarrow {\rm D}\varphi\in {\rm SO}(3)$.
		\subsection{The exponentiated tension-compression symmetric energy  (polyconvex for all $k>0$)}\label{exptc-energy}
	\begin{align}
	W(F)=\frac{\mu }{4\,k}\,\left( e^{k\,\left\|\frac{U}{(\det U)^{1/2}}-\left(\frac{U}{(\det U)^{1/2}}\right)\right\|^2}-1\right)+\frac{\mu+2\,\lambda }{4\, \widehat{k}}\,\left(e^{\widehat{k}(\log \det F)^2}-1\right).
	\end{align}

	\subsection{The Richter  energy}\label{Richter-energy}
	\begin{align}
W_{\rm Richter}(F)=\mu\det V\Big(\tr(V)-4\Big)-\mu, \qquad \lambda=0.	\end{align}

	\subsection{The one parameter energy convex in $C$ \cite{gao2015convexity} }\label{convex-energy}
	\begin{align}W(C)=\frac{\mu}{8} (\|C\|^2-2 \log \det C-3),\qquad \lambda=0. \end{align}
%\subsection{The Carroll  energy }\label{Carol-energy}
%	\begin{align}W_{\rm Carroll}(B)=&\,\frac{1}{12} (10 \mu -3 \lambda )\, I_1(B)+\frac{3 \lambda +2 \mu }{1296}\, I_1^4(B)-\sqrt{3} \mu\, \sqrt{I_2(B)}-\frac{3(3 \lambda +2 \mu )}{16},
%	\end{align}
%	where $I_1$ and $I_2$ are the first two invariants of the symmetric matrices.

	\subsection{The slightly compressible Carroll  energy \cite{carroll2011strain} with volumetric-isochoric split}\label{Carolvi-energy}
	\begin{align}W_{\rm sCarroll}(B)=\,\mu\, I_1(\frac{B}{(\det B)^{1/3}})&+\frac{1}{648} \left(\left(-\sqrt{3}\right) c-3 \mu \right)\, I_1^4(\frac{B}{(\det B)^{1/3}})+c\, \sqrt{I_2(\frac{B}{(\det B)^{1/3}})}\\&+\frac{\kappa}{8\,\widehat{k}}\left( e^{\widehat{k} (\log \det F)^2}-1\right)-\frac{7}{8} \left(\sqrt{3} c+3 \mu \right),\notag
	\end{align}
	  where $\mu>0, c>0$.
	  Since
	  \begin{align}
	  	I_1(\frac{B}{(\det B)^{1/3}})&=\langle \frac{B}{(\det B)^{1/3}},\id_ 3\rangle=\frac{1}{(\det F)^{2/3}}\|F\|^2,\\
	I_2(\frac{B}{(\det B)^{1/3}})&=\langle \Cof\left(\frac{B}{(\det B)^{1/3}}\right),\id_ 3\rangle=\det \left(\frac{B}{(\det B)^{1/3}}\right)\langle \left(\frac{B}{(\det B)^{1/3}}\right)^{-1},\id_ 3\rangle=\frac{1}{(\det F)^{4/3}}\|\Cof\, F\|^2,\notag
	  \end{align}
	  the Carroll energy reads
	  	\begin{align}W_{\rm sCarroll}(B)=\,\mu\, \frac{\|F\|^2}{(\det F)^{2/3}}&+\frac{1}{648} \left(\left(-\sqrt{3}\right) c-3 \mu \right)\, \frac{\|F\|^8}{(\det F)^{8/3}}+c\, \frac{\|\Cof\, F\|}{(\det F)^{2/3}}\\&+\frac{\kappa}{2\,\widehat{k}} \left( e^{\widehat{k} (\log \det F)^2}-1\right)-\frac{7}{8} \left(\sqrt{3} c+3 \mu \right), \qquad c>0.\notag
	  \end{align}

	  	Another rewriting  of the Carroll energy for slightly compressible materials is
	  		\begin{align}W_{\rm mCarroll}(B)=\,\mu\,\left(\underbrace{ \frac{\|F\|^2}{(\det F)^{2/3}}-3}_{\geq0}\right)&+\frac{1}{648} \left(\left(-\sqrt{3}\right) c-3 \mu \right)\,\left(\underbrace{ \left(\frac{\|F\|^2}{(\det F)^{2/3}}\right)^4-3^4}_{\geq 0}\right)\\&+c\, \left(\underbrace{\frac{\|\Cof\, F\|}{(\det F)^{2/3}}-\sqrt{3}}_{\geq 0}\right)+\frac{\kappa}{8\,\widehat{k}} \left( e^{\widehat{k} (\log \det F)^2}-1\right), \qquad c>0.\notag
	  	\end{align}
	  	Here, we have used that $\|F\|^3\geq 3\sqrt{3}(\det F)$ and  $\frac{\sqrt{x\, y+y \,z+z\,x}}{(x \,y\, z)^{2/3}}\geq \sqrt{3}$, $\forall x, y, z>0$.
	  	\subsection{The modified  slightly compressible Carroll  energy  with volumetric-isochoric split}\label{mCarolvi-energy}
	  	
	  	We propose 	a modified form of the Carroll energy, too, i.e.,
	  	\begin{align}W_{\rm mCarroll}(B)=&\,\mu\,\left( \frac{\|F\|^2}{(\det F)^{2/3}}-3\right)+b\,\left( \frac{\|F\|^2}{(\det F)^{2/3}}-3\right)^4\\&\quad\  -\sqrt{3} \,\mu\, \left(\frac{\|\Cof\, F\|}{(\det F)^{2/3}}-\sqrt{3}\right)+\frac{\kappa}{8\,\widehat{k}} e^{k (\log \det F)^2}, \qquad b>0.\notag
	  	\end{align}
	  \subsection{The  Gao energy}
	 A simple energy is proposed in \cite{gao1997large}
	 \begin{align}
	 	W_{\rm Gao}(C)&=a\, (I_1^\alpha(C)+I_{-1}^\alpha(C))-{2\cdot 3^\alpha a}=a\,([\tr(C)]^\alpha+[\tr(C^{-1})]^\alpha)-2\cdot 3^\alpha a\\
	 	&=a\,(\|F\|^\alpha+\|F^{-1}\|^\alpha)-2\cdot 3^\alpha a,
	 \end{align}
	 where $I_{-1}(C)=\tr(C^{-1})=\frac{I_2(C)}{I_3(C)}.$ For compatibility with linear elasticity $\alpha=\frac{3 \lambda+2 \mu}{2 \mu}$ and $a=\frac{\lambda}{8\ 3^{\alpha -2} \alpha  (\alpha +2)}$.
	
	  Therefore, the Gao energy is
	  \begin{align}
	  	W_{\rm Gao}(C)=\frac{\lambda}{8\ 3^{\alpha -2} \alpha  (\alpha +2)}\, (I_1^\frac{3 \lambda+2 \mu}{2 \mu}(C)+I_{-1}^\frac{3 \lambda+2 \mu}{2 \mu}(C))-\frac{18\,\lambda}{8 \alpha  (\alpha +2)}, \qquad \lambda>0.
	  \end{align}
	  \subsection{The Murnaghan energy}
	  An energy often used in nonlinear acoustoelasticity is proposed in \cite{murnaghan1937finite}
	  \begin{align}
	  	W_{\rm Murnaghan}(E)=\frac{1}{4}\,\left[\frac{1}{2}(\lambda+2\mu)I_1^2(E)-2\, \mu\, I_2(E)+\frac{1}{3}(l+2m)I_1^3(E)-2\, m \, I_1(E)I_2(E)+n\, I_3(E)\right],
	  \end{align}
	  where $l, m, n$ are the Murnaghan third-order elastic moduli (tabulated for many materials) and $E=\frac{1}{2}(C-\id_3)$.
	  
	  Another form used in the literature is 
	   \begin{align}
	  	W_{\rm Murnaghan}(E)=\mu \,\tr E^2+\frac{\lambda}{2}(\tr E)^2+\frac{\nu_1}{6}(\tr E)^3+\nu_2 (\tr E)(\tr E^2)+\frac{4\,\nu_3}{3}\tr E^3,
	  	\end{align}
where $\nu_1, \nu_2, \nu_3$ are the  Murnaghan third order elasticity constants satisfying
\begin{align}
	l=\frac{\nu_1}{2}+\nu_2, \qquad m=\nu_2+2\, \nu_3, \qquad n=4\, \nu_3.
	\end{align}
	In this form, the Murnaghan energy is the most general isotropic formulation up to order three in the strain $E$.
\section{Volumetric terms}\setcounterpageref{equation}{0}

Different expressions of the volumetric part (the term depending on $J=\det F$)  may lead to different constitutive properties.
	There are many possibilities to choose  such a function  in $\det F$, see \cite{Hartmann_Neff02}, e.g.,
\begin{align}
	h:[0,\infty)\to \mathbb{R}, \ h(J)=\begin{cases}
		(\log  J)^2, &\text{not convex, symmetric for } J\mapsto 1/J ,\\
		\frac{1}{2\widehat{k}}e^{\widehat{k}(\log\,  J)^2}, &\text{convex for} \ \ \widehat{k}>3/8, \text{ symmetric for } J\mapsto 1/J\vspace{1.2mm}\\
		\frac{1}{4}(J-\frac{1}{J})^2&\text{convex  and } h'(1)=2, \text{ symmetric for } J\mapsto 1/J\vspace{1.2mm}\\
		\frac{1}{4\widehat{k}}	e^{\widehat{k}(J-\frac{1}{J})^2}, &\text{convex for} \ \ \widehat{k}>0, \text{ symmetric for } J\mapsto 1/J\vspace{1.2mm}\\
		J^2-\log  \, J&\text{convex}, \text{ not symmetric for } J\mapsto 1/J\vspace{1.2mm}\\
		\frac{1}{2}\left[\frac{J^2-1}{2}-\log \, J\right] &\text{convex}, \text{ not symmetric for } J\mapsto 1/J\\
		-1/2\log \left[1 - \frac{J^2 + 1/J^2 - 2}{10}\right], & \sqrt{6-\sqrt{35}}<J<\sqrt{6+\sqrt{35}}, \text{ symmetric for } J\mapsto 1/J
	\end{cases}
\end{align}

For $h:\mathbb{R}_+\to \mathbb{R}$ convex,  rank-one convexity/polyconvexity and TSTS-M$^+$ are both satisfied (see, e.g., Section \ref{ips2}) for the volumetric term alone.

If, on the other hand \begin{align}h''(J)\, J+h'(J)>0,\end{align} then Hill's inequality is satisfied, see  \cite{korobeynikov2025two}. To see this, define
 $\widehat{h}:\mathbb{R}\to \mathbb{R}$ by $\widehat{h}(\log J)=h(J)$, for all $J\in \mathbb{R}_+$. Then, since $ h'(J)=\widehat{h}'(\log J)\, J^{-1}$, \ $ h''(J)=\widehat{h}''(\log J)\, J^{-2}-\widehat{h}'(\log J)\, J^{-2}$ and $\widehat{h}''(\log J)\, J^{-2}=h''(J)+h'(J)\, J^{-1}$, the  Hill's inequality is satisfied if and only if $\widehat{h}$ is strictly convex in $\xi=\log \det F$, i.e., $\widehat{h}''(\xi)>0$ for all $\xi>0$.
	\section{Constitutive requirements in nonlinear elasticity}\setcounter{equation}{0}\label{ci}
	\subsection{Invertibility for compressible materials }\label{ips}

	We consider the  general isotropic constitutive equation
	\begin{equation}
		\Sigma=\Sigma(U)\qquad \qquad \Sigma:{\rm Sym}^{++}(3)\to \Sym(3)
	\end{equation}
	where $\Sigma$ is some symmetric stress tensor and $U\in{\rm Sym}^{++}(3)$ is the stretch tensor. 	Since, in general, it is not easy to work with tensors (matrices) in three dimensions, we consider the singular values (the principal stretches) $\lambda_1$, $\lambda_2$, $\lambda_3$ of $F$, i.e., the positive eigenvalues  of $U=\sqrt{F^TF}$.
	If $\Sigma_f\col\Symn\to\Symn$ is an	 isotropic tensor function satisfying \begin{align}\Sigma_f(Q^T\cdot\diag(\lambda_1,\lambda_2,\lambda_3)\cdot Q)=Q^T\cdot\Sigma_f(\diag(\lambda_1,\lambda_2,\lambda_3))\cdot Q\end{align} then
	\begin{align}
		\Sigma_f(U):=\Sigma_f(\underbrace{Q^T\cdot\diag(\lambda_1,\lambda_2,\lambda_3)\cdot Q}_{S\,\in\,\Sym(3)}) = \underbrace{Q^T\cdot\diag(f(\lambda_1,\lambda_2,\lambda_3))\cdot Q}_{\Sigma_f(S)\,\in\,\Sym(3)} \quad\forall\,Q\in\On\label{eq:introductionMatrixFunction}
	\end{align}
	with a vector-function $f=(f_1,f_2,f_3)\col\R^3\to\R^3$ which fulfils
	\[
	f_i(\lambda_{\pi(1)},\lambda_{\pi(2)},\lambda_{\pi(3)}) = f_{\pi(i)}(\lambda_1,\lambda_2,\lambda_3)
	\]
	for any permutation $\pi\col\{1,2,3\}\to\{1,2,3\}$.

	The functions $f$ and $\Sigma_f$ related by eq.~\eqref{eq:introductionMatrixFunction} share a number of properties related to invertibility and monotonicity \cite{ghiba2024biot}.
	\begin{proposition}\label{invprop}
		Assume that $\widetilde{\Sigma}_{\widetilde{f}}:{\rm Sym}(3)\to{\rm Sym}(3)$ is an isotropic $C^1$-function defined by the vector-function ${\widetilde{f}}:\mathbb{R}^3\to \mathbb{R}^3$ such that \begin{enumerate}
			\item ${\rm D}{\widetilde{f}}\,(x_1,x_2,x_3)$ is invertible for any $(x_1,x_2,x_3)\in \mathbb{R}^3$;
			\item $\|{\widetilde{f}}\,(x_1,x_2,x_3)\|_{\mathbb{R}^3}\to \infty$ as $\|(x_1,x_2,x_3)\|_{\mathbb{R}^3}\to \infty$.
			
		\end{enumerate}
		Then $U\mapsto \widetilde{\Sigma}_{\widetilde{f}}(U)$ is a global diffeomorphism from ${\rm Sym}(3)$ to ${\rm Sym}(3)$.
	\end{proposition}

	\begin{corollary}\label{corinvprop1}
		Assume that $\Sigma_f:{\rm Sym}^{++}(3)\to{\rm Sym}(3)$ is an isotropic $C^1$-function such that\begin{enumerate}
			\item ${\rm D}f\,(\lambda_1,\lambda_2,\lambda_3)$  is invertible for any $(\lambda_1,\lambda_2,\lambda_3)\in \mathbb{R}^3_+$;
			\item $\|f(\lambda_1,\lambda_2,\lambda_3)\|_{\mathbb{R}^3}\to \infty$ as  $\|(\log\lambda_1,\log\lambda_2,\log\lambda_3)\|_{\mathbb{R}^3}\to \infty$.
		\end{enumerate}
		Then $U\mapsto \Sigma_f(U)$ is a global diffeomorphism from ${\rm Sym}^{++}(3)$ to ${\rm Sym}(3)$.
	\end{corollary}

	\subsection{Hilbert-monotonicity }\label{ips2}
	
	For our purposes, we now recall some related notions of monotonicity.

	\begin{definition}\cite{NeffMartin14}
		A tensor function $\Sigma_f\col{\rm Sym}^{++}(3)\to\Symn\,$ is called  \textbf{strictly  Hilbert-monotone} if
		\begin{align}
			\iprod{\Sigma_f(U)-\Sigma_f(\overline U),\,U-\overline U}_{\R^{3\times 3}}>0\qquad\forall\,U\neq\overline U\in{\rm Sym}^{++}(3)\,.\label{eq:introductionMatrixMonotonicity}
		\end{align}
		We refer to this inequality as \textbf{strict Hilbert-space matrix-monotonicity} of the tensor function $\Sigma_f$. A tensor function $\Sigma_f\col{\rm Sym}^{++}(3)\to\Symn\,$ is called  \textbf{Hilbert-monotone} if
		\begin{align}
			\iprod{\Sigma_f(U)-\Sigma_f(\overline U),\,U-\overline U}_{\R^{3\times 3}}\geq 0\qquad\forall\,U,\,\overline U\in{\rm Sym}^{++}(3)\,.\label{eq:introductionMatrixMonotonicity1}
		\end{align}
	\end{definition}
	\begin{definition}\cite{NeffMartin14}
		A vector function  $f\col\R^3_+\to\R^3$ is \textbf{strictly vector monotone} if
		\begin{align}
			 \iprod{f(\lambda)-f(\overline\lambda),\,\lambda-\overline\lambda}_{\R^3}>0\qquad\forall\lambda\neq\overline\lambda\in\R^3_+,
		\end{align}
		and it is is   \textbf{vector monotone} if
		\begin{align}
			\iprod{f(\lambda)-f(\overline\lambda),\,\lambda-\overline\lambda}_{\R^3}\geq 0\qquad\forall\lambda,\lambda\in\R^3_+,
		\end{align}
	\end{definition}
	\begin{definition}
		A continuously differentiable tensor function $\Sigma_f\col{\rm Sym}^{++}(3)\to\Symn\,$ is called  \textbf{strongly Hilbert-monotone} if
		\[
		\iprod{{\rm D}\Sigma_f(U).H,H} > 0 \qquad\text{for all }\;{U\in\rm Sym}^{++}(3),\;H\in{\rm Sym}(3)
		\,.
		\]
		%	\begin{align}
			%	\iprod{\Sigma_f(U)-\Sigma_f(\overline U),\,U-\overline U}_{\R^{3\times 3}}\geq \|U-\overline U\|_{\R^{3\times 3}}\qquad\forall\,U,\overline U\in{\rm Sym}^{++}(3)\,.\label{eq:introductionMatrixMonotonicity2}
			%	\end{align}
	\end{definition}
	
	\begin{definition}\cite{NeffMartin14}
		A continuously differentiable vector function  $f\col\R^3_+\to\R^3$ is called \textbf{strongly vector monotone} if
		\[
		\iprod{{\rm D}f(\lambda).h,h} > 0 \qquad\text{for all }\;\lambda\in\R^3_+,\;h\in\R^3
		\,\Longleftrightarrow\ \iprod{\sym\,{\rm D}f(\lambda).h,h} > 0 \qquad\text{for all }\;\lambda\in\R^3_+,\quad h\in\R^3.
		\]
		%	\begin{align}
			%	\iprod{f(\lambda)-f(\overline\lambda),\,\lambda-\overline\lambda}_{\R^3}\geq \|\lambda-\overline\lambda\|_{\R^3}\qquad\forall\lambda\neq\overline\lambda\in\R^3_+.
			%	\end{align}
	\end{definition}

	%For a continuously differentiable function $f$ on a convex set, where the derivative ${\rm D}f$ is self-adjoint and invertible throughout, the positive definiteness of ${\rm D}f$ at a single point is enough to guarantee that $f$ is strictly monotone across the entire set.

	\noindent In a forthcoming paper \cite{MartinVossGhibaNeff}, we discuss the following result, thereby expanding on Ogden's work \cite[last page in the Appendix]{Ogden83}, following Hill's seminal contributions  \cite{hill1968constitutivea,hill1968constitutiveb,hill1970constitutive}:
	\begin{theorem}
		\label{theorem:mainResult}
		A sufficiently regular symmetric function $f\col\R^3_+\to\R^3$ is (strictly/strongly) vector-monotone if and only if $\Sigma_f$ is (strictly/strongly) matrix-monotone.
	\end{theorem}
	\noindent Hence, in view of Theorem \ref{theorem:mainResult}, the following  holds true for hyperelasticity, assuming sufficient regularity:
	\begin{align*}
		U\mapsto T_{\rm Biot}(U) \ \ \text{is Hilbert-monotone}\quad &\Longleftrightarrow\quad (\lambda_1,\lambda_2,\lambda_3)\mapsto \widehat{T}(\lambda_1,\lambda_2, \lambda_3) \ \ \text{is vector monotone}\notag\\&\Longleftrightarrow\quad {\rm D}\widehat{T}(\lambda_1,\lambda_2, \lambda_3)\in {\rm Sym}^{+}(3)\quad\forall(\lambda_1,\lambda_2, \lambda_3)\in\mathbb{R}_+^3,
		\\[.7em]
		%\end{align*}
		%and
		%\begin{align}
		U\mapsto T_{\rm Biot}(U) \ \ \text{is strictly Hilbert-monotone}\quad &\Longleftrightarrow\quad (\lambda_1,\lambda_2,\lambda_3)\mapsto \widehat{T}(\lambda_1,\lambda_2, \lambda_3) \ \ \text{is strictly vector monotone,}\notag
		\\[.7em]
		U\mapsto T_{\rm Biot}(U) \ \ \text{is strongly Hilbert-monotone}\quad &\Longleftrightarrow\quad (\lambda_1,\lambda_2,\lambda_3)\mapsto \widehat{T}(\lambda_1,\lambda_2, \lambda_3) \ \ \text{is strongly vector monotone}\notag\\&\Longleftrightarrow\quad {\rm D}\widehat{T}(\lambda_1,\lambda_2, \lambda_3)\in {\rm Sym}^{++}(3)\quad\forall(\lambda_1,\lambda_2, \lambda_3)\in\mathbb{R}_+^3.
		%\\&\Longrightarrow\quad {\rm D}\widehat{T}(\lambda_1,\lambda_2, \lambda_3)\in {\rm Sym}^{+}(3).
	\end{align*}
	Note that
	\begin{enumerate}
		\item $U\mapsto T_{\rm Biot}(U)$ monotone is {\it Krawietz $M$-condition} \cite{krawietz1975comprehensive} $\Leftrightarrow$ \ $U\mapsto \widetilde{W}(U)$ is convex.
		\item $\log V\mapsto \widehat{\tau}(\log V)$ monotone is {\it Hill's inequality} \cite{hill1968constitutivea,hill1968constitutiveb,ogden1977inequalities} $\Leftrightarrow$ \ $\log V\mapsto \widehat{W}(\log V)$ is convex.
		\item $\log V\mapsto \widehat{\sigma}(\log V)$ monotone is equivalent to  the new {\it corotational stability postulate} (CSP) \cite{d2024constitutive,neff2024hypo,neff2024natural} and coincides with {TSTS-M}$^{++}$ \cite{jog2013conditions,NeffGhibaLankeit,neff2024hypo}.
	\end{enumerate}

The {TSTS-M}$^{++}$ condition may be instrumental in obtaining existence results for nonlinear elastic materials \cite{neff2025rateI,neff2025rateII}. The {TSTS-M}$^{++}$ condition has been shown to be equivalent to the corotational stability postulate \cite{neff2024hypo}.  Moreover, {TSTS-M}$^{++}$ implies monotone increasing Cauchy stresses in standard homogeneous tests \cite{neff2025corotational} in which the principle axis do not rotate (excluding, e.g., simple shear).

 In view of the foregoing, the last author has offered a 500 \euro \ challenge to find an isotropic hyperelastic energy that satisfies LH-ellipticity (polyconvexity) and {TSTS-M}$^{++}$ everywhere on ${\rm GL}^+(3)$ \cite{NeffIutam2024}.

\subsection{On the difference between local and global invertibility}

Since for invertibility we only check the  local condition
\begin{align}
\det \D_\lambda\sigma (\lambda)\neq 0 \ \ \  \Leftrightarrow\ \ \  	\det \D_V \sigma (V)\neq 0 \ \ \ \Leftrightarrow\ \ \   	\det \D_{\log V} \widehat{\sigma} (\log V)\neq 0\ \ \Leftrightarrow\ \ 	\ \det \D_{\log \lambda} \widehat{\sigma} (\log\lambda)\neq 0
	\end{align}
	it is useful to remember the limitations of such a check.
	
\begin{figure}[!h]
		\centering
		\begin{minipage}{.25\textwidth}\centering
			\includegraphics[scale=0.4]{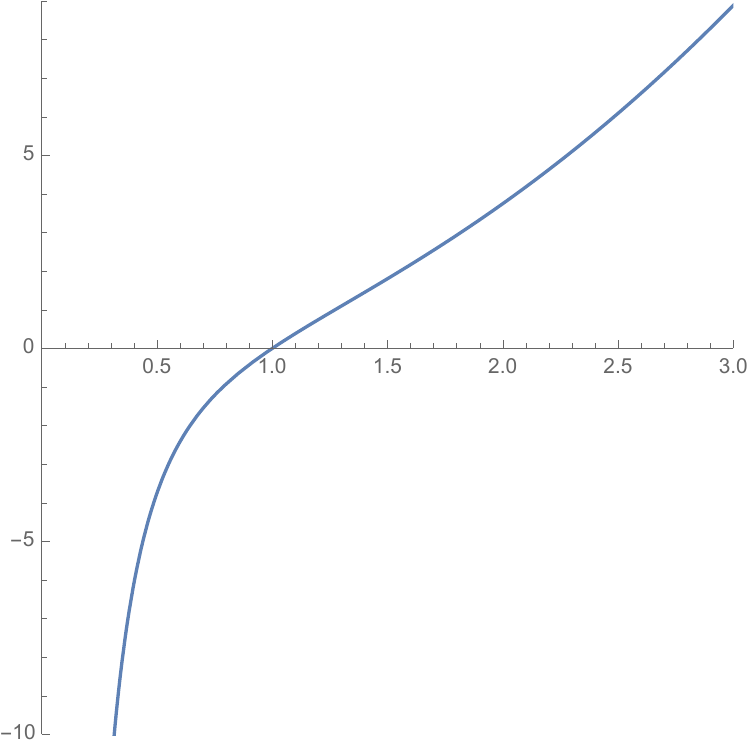}
			\caption{\footnotesize{Local vs. global invertibility for  $\lambda>0$: $\D \sigma_1(\lambda )\neq 0$  and we have global invertibility. }}
			\label{figP1}
		\end{minipage}
		\qquad \quad
		\begin{minipage}{.25\textwidth}\centering
			\includegraphics[scale=0.4]{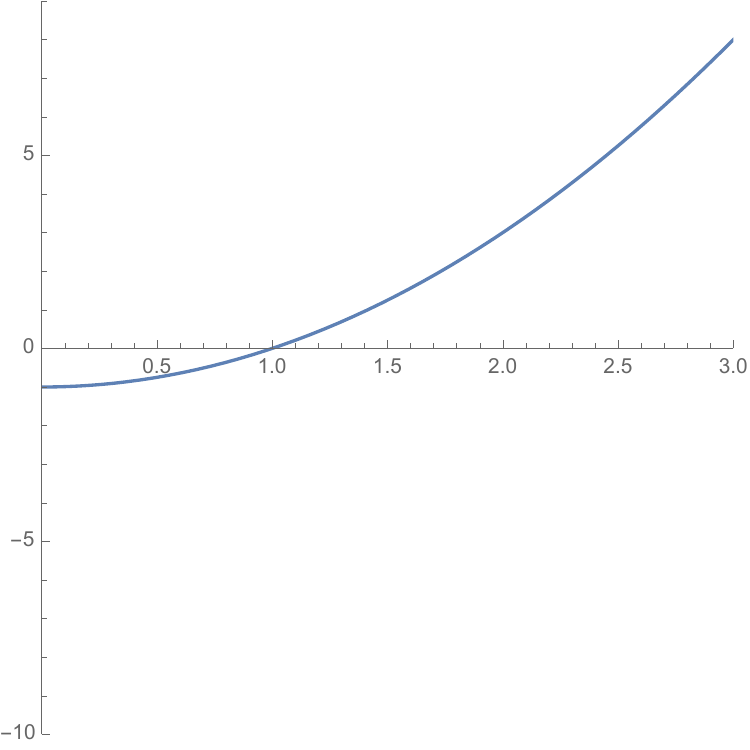}
			\caption{\footnotesize{ Local vs. global invertibility for $\lambda>0$: $\D \sigma_2(\lambda )\neq 0$ but it is   not globally invertible.}}
			\label{figP2}
					\end{minipage}
				\qquad \quad
			\begin{minipage}{.25\textwidth}\centering
				\includegraphics[scale=0.4]{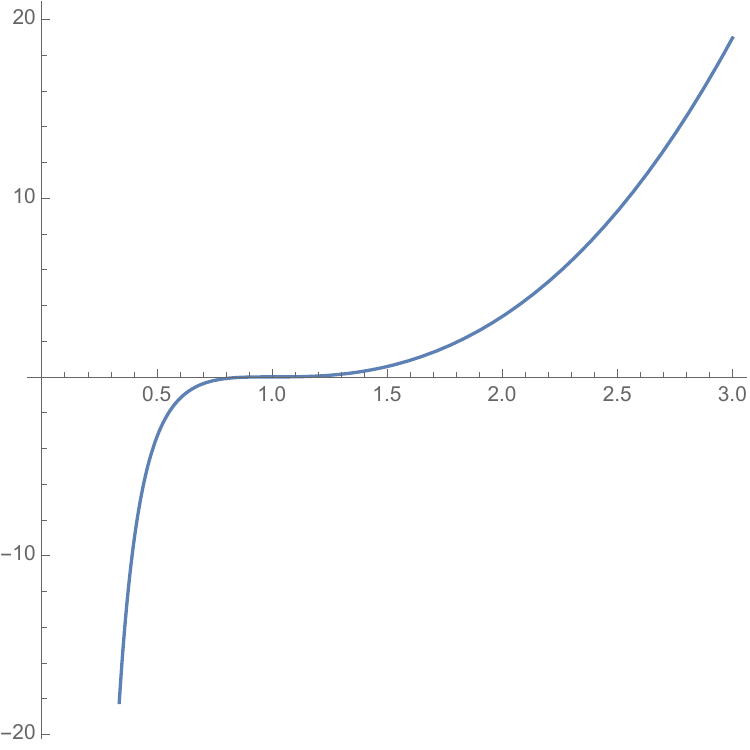}
				\caption{\footnotesize{Global invertibility does not  necessarily imply $\D \sigma_3(\lambda )\neq 0$ for $\lambda>0$.}}
				\label{figP3}
		\end{minipage}
		
	\end{figure}

	Consider $\sigma_1:\mathbb{R}_+\to \mathbb{R}$, $\sigma_1(\lambda)=\frac{\mu}{4}\left(\lambda^2-\frac{1}{\lambda^2}\right)$  versus $\sigma_2:\mathbb{R}_+\to \mathbb{R}$, $\sigma_2(\lambda)=\frac{\mu}{4}\left(\lambda^2-1\right)$. Both satisfy local invertibility, but only the first example is also globally invertible, see Figure \ref{figP1} and \ref{figP2}.
	
	Furthermore, we need to remember that global invertibility does not imply $\D_{\lambda}\sigma(\lambda)\neq 0$, see the one-dimensional example $\sigma_3:\mathbb{R}_+\to \mathbb{R}$, $\sigma_3(\lambda)=(\lambda-\frac{1}{\lambda})^3$, see Figure \ref{figP3}.
	
		\restylefloat*{figure}
	
\section{The method of study}

For the family of energies listed in the next section, we implement numerically the check of the hypothesis from the results presented above and  in Section \ref{ci}. For instance, for the energy considered before in the introduction section just as an example, we check the monotonicity and the invertibility of some  stress-strain relations using the results established in terms of the principal stresses and of principal stretches. The regions from  Figures \ref{1}-\ref{12} are those on which the local monotonicity and the local invertibility are  satisfies numerically.
There are also some other	candidates for the part of the energy  depending on $\det F$ (the  purely volumetric part). The properties of the convex functions $h$ in $\det F$ are important and influences the local invertibility and local monotonicity.

However, in order to check the regions on which the local monotonicity and the local invertibility are  properly satisfied, we do not use some already implemented function in some math software, but we have implemented our own  numerical variants. This is necessary because we have identified some problems which can arise if one uses directly the already implemented functions. One simple example useful to understand the problems is that most functions graphically represent the domain where certain conditions are satisfied, but it is not possible to see if there are subdomains inside these regions where these conditions are not satisfied. Apart from this somewhat trivial observation, since we are working with expressions that quickly lead to very large numbers, other problems may arise, as we will explain below.

Specifically, the computations involve evaluating symbolic expressions at a series of points in the domain of interest.
One of the variants consists of using the RegionPlot3D function in Mathematica to generate graphical representations that help identify whether there are points or areas where the hypotheses are not verified.
For this method, a series of issues arise that make the provided result not always "certain". This is determined, for instance,  by  at least one of the following reasons (the examples are extracted from the computations we have seen on our numerical simulations):
\begin{enumerate}
	
	\item Overflow. Using the following Mathematica function
	
	$RegionPlot3D[ Evaluate[-Exp[10^c] > 0  ], \{a, -2, 2\}, \{b, -1, 1\}, \{c, -20, 20\},  WorkingPrecision -> 10]$
	
	we observe in Figure \ref{Overflow} that a region is indicated where the exponential function is negative, which is clearly incorrect. Indeed, Mathematica gives us a warning overflow message and we have to take it into account and having doubts about the graphical representation provided as output.
	\begin{figure}[h!]
		\centering
		\begin{minipage}{.47\textwidth}\centering
			\includegraphics[scale=0.15]{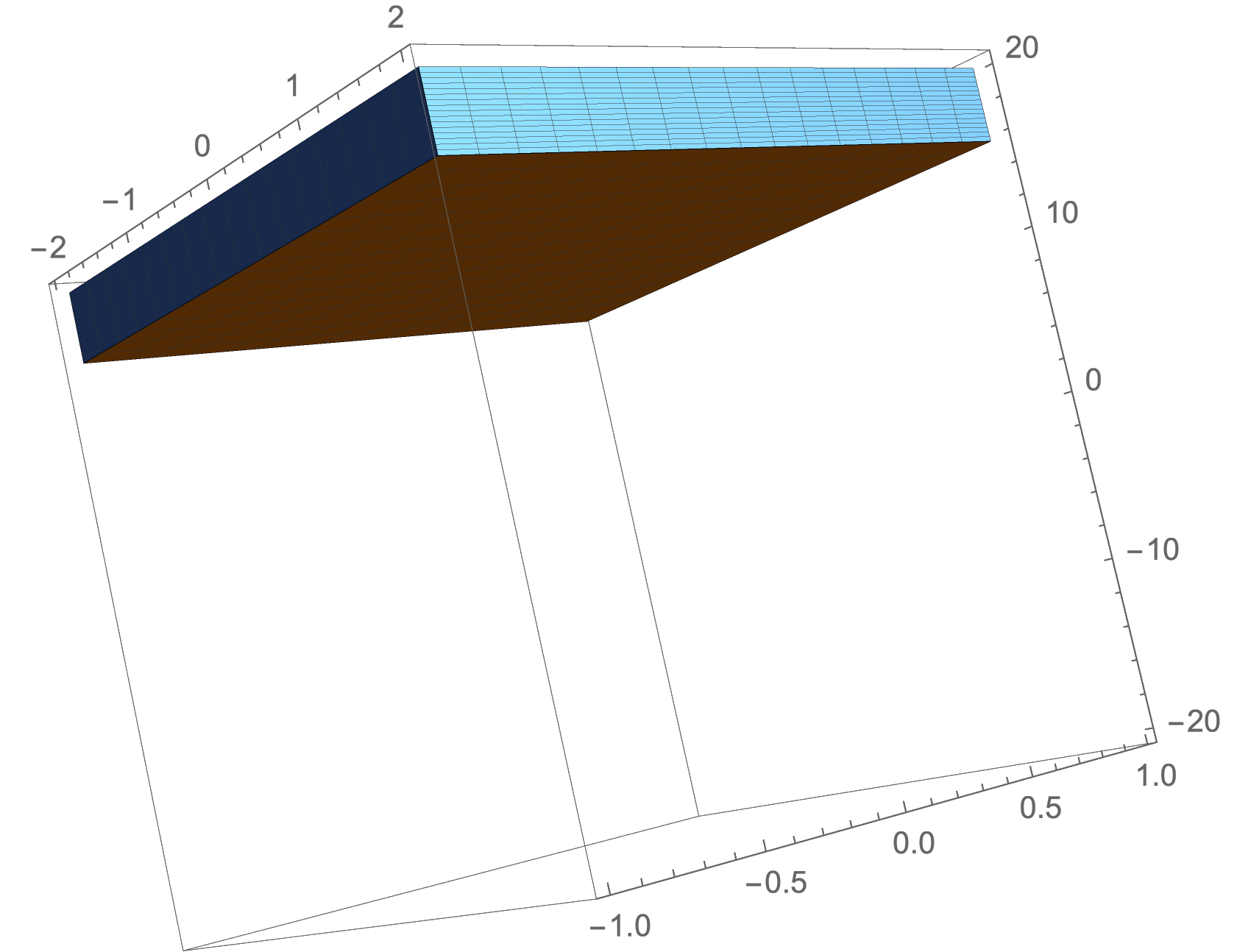}
			\caption{\footnotesize{ Example of overflow}}
			\label{Overflow}
		\end{minipage}
		\qquad 	
		\begin{minipage}{.47\textwidth}\centering
			\includegraphics[scale=0.15]{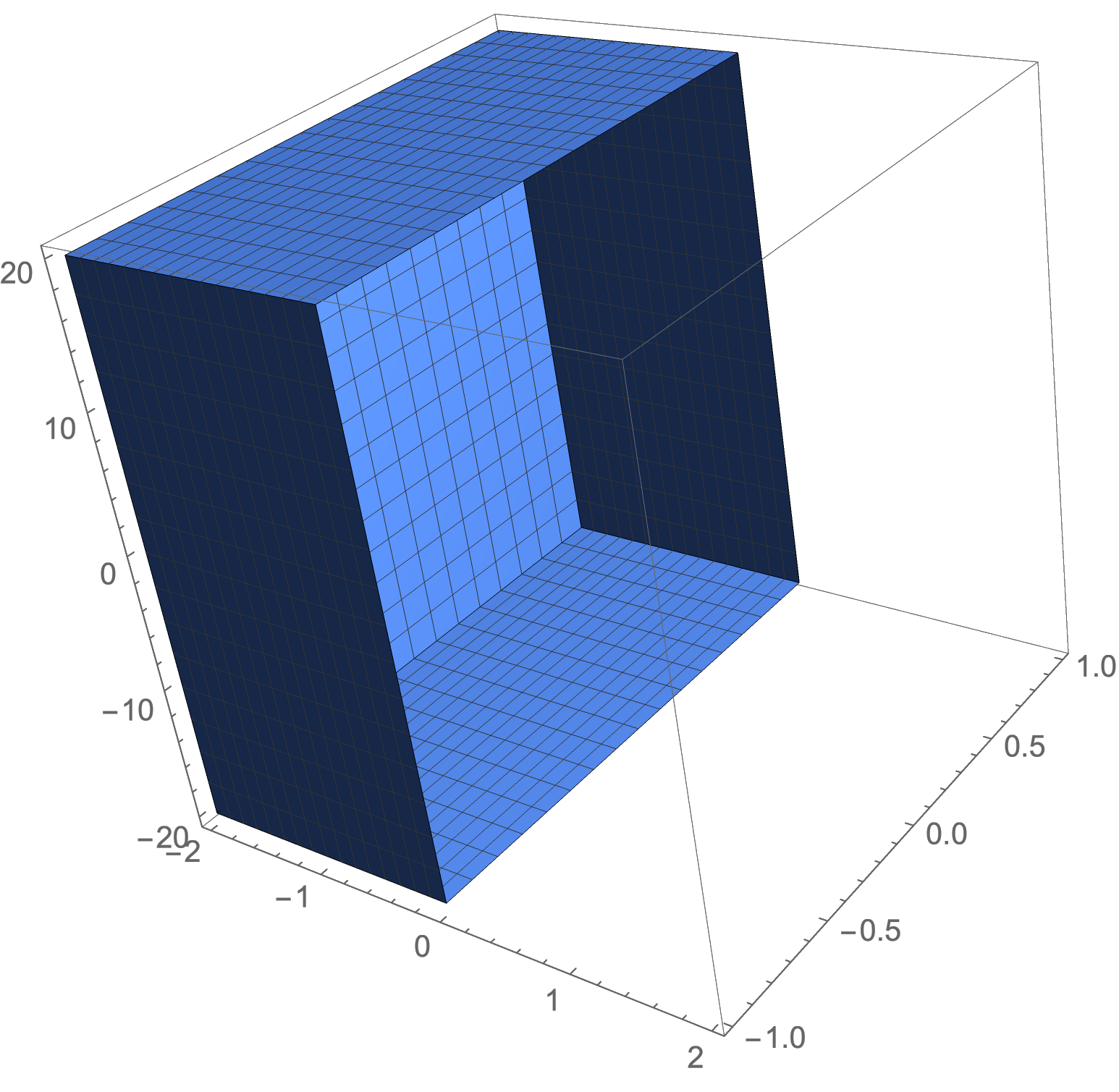}
			\caption{\footnotesize{ Example of indeterminate}}
			\label{Indeterminate}
		\end{minipage}\qquad
	\end{figure}

	\item Indeterminate. Using the following Mathematica function
	
	$RegionPlot3D[ Evaluate[Exp[1/(1 - Exp[500*a]/(1 + Exp[500*a]))] > 0  ], \{a, -2,  2\}, \{b, -1, 1\}, \{c, -20, 20\},$\qquad $ WorkingPrecision -> 10]
	$
	
	we observe in Figure \ref{Indeterminate} that a region is indicated where the considered function is positive, which is clearly incorrect. Indeed, Mathematica gives us a warning about an indeterminate expression, and, again,   we have to take it into account and not to misinterpret the results.

	\item Badly conditioned matrix
	
	A badly conditioned matrix is one where small changes in the input can lead to large changes in the output when solving equations or performing computations. This happens because the matrix has a high condition number, indicating numerical instability. Such matrices can cause significant errors in calculations and they arise in our numerical simulations if we are not careful.
	In the following example, it can be observed that for the matrix $M = \left(
	\begin{array}{ccc}
		1 & 2 & 3.2 \\
		1 & 2 & 3.2 \\
		2 & 1.1 & 1.2 \\
	\end{array}
	\right)$ we obtain results with different signs for $\det(M)$ and $\det(10^{22}M)$, i.e., $\det(M)=5.15143\times 10^{-16}$ and $\det(10^{22}M)=-1.64846\times 10^{51}$.
	
\end{enumerate}

To avoid the issues mentioned earlier as far as possible, a second method for numerical evaluation of symbolic expressions was implemented. The main goal was to obtain much finer control over the points where the obtained values fall into one of the cases presented above. Since a negative value can invalidate properties like local monotonicity or local invertibility, avoiding false negative evaluations is critical. With this in mind, a second approach was implemented, using the $C\#$ language for implementation.

The first step was to implement the \emph{BigNumber} class, which corresponds to elements of the form $a*10^b$, where $a$ and $b$ are of type \emph{Decimal}. The maximum value for the \emph{Decimal} type in $C\#$ is 79228162514264337593543950335, with 29 decimal places (represented on 16 bytes). In this class, the operators corresponding to arithmetic operations have been overloaded and elementary functions have been implemented. Focus has been placed on an optimized implementation of these functions, ensuring that exceptions are generated if the maximum/minimum range of values is exceeded, thus providing very fine-grained control over how the result of the expression evaluation should be interpreted.

An alternative approach was also implemented, using the BigRational type (from the library with the same name) instead of Decimal. However, in the tests performed by us,  no significant decrease is obersed in the number of points where out-of-range exceptions were generated. Since the execution time increased significantly, the use of BigRational was abandoned, and the $a*10^b$ format, where $a$ and $b$ are Decimal types, was retained.

The second step was to implement the modelling of symbolic expressions in a tree structure. Each node of this tree corresponds to an arithmetic operation or an elementary function. The classes corresponding to these operations inherit from the abstract class \emph{Expression}, which has the abstract method:

\texttt{	public abstract BigNumber Val \{ get; \}}

\begin{figure}[h!]
	\centering
	\includegraphics[scale=0.35]{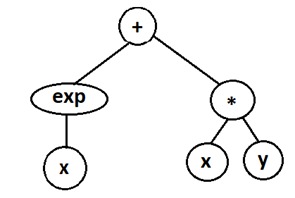}
	\caption{A tree strategy for evaluation of a symbolic expression.}
	\label{Tree}
\end{figure}

The value of an expression is obtained by reading this value for the root node (the reading being performed for each of the points of interest).
The image below, Figure \ref{Tree}, represents the tree corresponding to the symbolic expression: $e^x +xy$

For most of the energies considered, the symbolic expressions corresponding to the hypotheses to be tested are complex (some exceeding 90,000 characters).

By building the corresponding tree we have control at the level of each of its nodes. Thus, to reduce the probability of an incorrect return value, whenever an out-of-range condition could occur, an exception is throwned, which is then catched and interpreted correctly.

A special attention was paid to \emph{Add} and \emph{Multi} type nodes corresponding to expressions of the form $\sum a_i - \sum b_j$ and $\prod a_i/\prod b_j$, respectively, optimizing the evaluation method, in order to minimize the risk of out of range or indeterminacy exceptions.

The next step was to implement the selection of points in the domain of interest. Two approaches were considered:

- based on a uniform distribution, specifying for each axis a minimum, a maximum and the number of elements,

- based on a normal distribution, specifying for each axis the mean, standard deviation and the number of elements.

For the results presented below, we used a set of $121^3$ elements $(x, y, z)$ from the domain $(0.2000]\times(0.2000]\times(0.2000]$ with
$x$, $y$, $z$ $\in \{ a_{i-1}+(a_i-a_{i-1})*k/10$| where $i=\overline{1,2,...,12}$, $k=\overline{0, 1,..., 10}\}$
and $\{a_i\}_{i=\{0,...,12\}} = \{ 10^{-8}, 10^{-5}, 10^{-2}, 10^{-1}, 1, 2, 10, 25, 50, 100, 500, 1000, 2000 \}$

An important aspect of the implemented algorithm is that when evaluating an expression for a data series, if an exception is detected (for example, an out-of-range calculation) or the value obtained invalidates the assumption conditions, an additional data set is automatically generated around that point. This additional set consists of points very close to the original one (within the definition range), and the evaluation is also performed on these points.

- If for a point  $(x_0, y_0, z_0)$ we catch an exception, then the conditions will be evaluated in another 216 points  from a neighborhood of the point, i.e., in $(x_0*a, y_0*b, z_0*c)$ where $a$,$b$,$c \in \{ (95+2k)/100| k=0,1,..,5\}$. If for every point from this neighborhood there is detected an exception then the point $(x_0, y_0, z_0)$ is marked as an uncertainty pointis.

-  If for a point $(x_0, y_0, z_0)$ we obtain a negative value, then the point is marked as a fail point; In this case  the conditions will be evaluated in another  300 points around the point will be evaluated. These points are uniformly distributed, 150 of them being at a distance of $10^{-8}$ and another 150 points at a distance of $10^{-7}$. In the implementation, the distances were calculated using the infinite norm and the fact that the points have positive coordinates was taken into account. Based on the number of points from this dataset for which negative values were obtained, we calculate the probability that the value at point $(x_0, y_0, z_0)$ is negative. The value, the point and the probability for the point with high probability are retained. The search stops if we find a point for which the evaluation of the expression is negative with probability close to 1.

At the end of the test the following are possible, see Tables \ref{tabelbc1}--\ref{tabelbc3}

- we found a point in which the value of the expression is negative with probability p; In this case we say that the energy fails the test with probability $p$.

- we do not find points with negative value and in this case we say that the energy can pass the test with a probability determined by the number of non uncertainty points and the total number of points.

Taking into account the above, in the table below, the meaning of the symbols is as follows:

$\checkmark\checkmark$ - it is already known analytically that the condition is satisfied \cite{korobeynikov2025two,NeffGhibaLankeit,neff2025rateI,neff2025rateII,neff2024hypo,neff2025corotational},

$\times\times$ - it is already known analytically that the condition is not satisfied  \cite{korobeynikov2025two,NeffGhibaLankeit,neff2025rateI,neff2025rateII,neff2024hypo,neff2025corotational},

$\checkmark$ - no points marked as counter-example point were found, and uncertainty points (calculus errors of the type given above) is below $5\%$,

$(\checkmark)$ - no points marked as counter-example point were found, and uncertainty points  (calculus errors of the type given above) is below $33\%$,

$((\checkmark))$ - no points marked as counter-example point were found, and uncertainty points  (calculus errors of the type given above) is below $66\%$,

$(?)$ - no points marked as counter-example point were found and uncertainty points  (calculus errors of the type given above) is above $66\%$,

$\times$ - a counter-example  point was found,

$(\times)$ - a counter-example point was found, but in a small neighborhood the values of the involved expressions are not stable (values fluctuate between very high and very low values).

	\section{Tables of the properties for the family of isotropic energies}

In this section we present our numerical evidences for or against local invertibility and local monotonicity of the stress-strain relations for various energies.

The tables below present the results obtained for the energies considered.

\begin{remark}
	For each given constitutive choice it is, in principle, possible to rigorously proof the statements of the numerical table. Here, the statements of the table indicate what can be hoped for.
\end{remark}

\begin{footnotesize}
	\begin{table}[hbt!]
		\centering
		\resizebox{16.5cm}{!}{ 	\begin{tabular} [hbt!]{|l|c|c|c|c|c|c|c|c|c|c|c|c|}
				\hline
				\footnotesize	\begin{minipage}{3cm}\footnotesize \vspace{3mm}\textbf{Energy}\vspace{3mm}\end{minipage}& \begin{minipage}{1cm} \footnotesize\textbf{$\sigma$-$V$ inv} \end{minipage}&\begin{minipage}{1.2cm} \footnotesize\textbf{$T_{\rm Biot}$-$U$ inv} \end{minipage} &\begin{minipage}{1cm} \footnotesize\textbf{$\tau$-$V$ inv} \end{minipage}& \begin{minipage}{1cm} \footnotesize\textbf{$\sigma$-$\log V$ inv} \end{minipage}&\begin{minipage}{1.2cm} \footnotesize\vspace{3mm}\textbf{$T_{\rm Biot}$-$\log U$ inv} \vspace{3mm}\end{minipage} &\begin{minipage}{1cm} \footnotesize\textbf{$\tau$-$\log V$ inv.} \end{minipage}
				& \begin{minipage}{1cm} \footnotesize\textbf{$\sigma$-$V$ mon} \end{minipage}&\begin{minipage}{1.2cm} \footnotesize\textbf{$T_{\rm Biot}$-$U$ mon} \end{minipage} &\begin{minipage}{1cm} \footnotesize\textbf{$\tau$-$V$ mon} \end{minipage}& \begin{minipage}{2cm} \footnotesize\textbf{$\sigma$-$\log V$ mon \\ (TSTS-M$^{++}$)} \end{minipage}&\begin{minipage}{1.2cm} \footnotesize\textbf{$T_{\rm Biot}$ -$\log U$ mon} \end{minipage} &\begin{minipage}{1.5cm} \footnotesize\textbf{$\tau$-$\log V$ mon\\ Hill's inequality} \end{minipage}
				\\\hline \begin{minipage}{3.1cm}
					\footnotesize	
					\vspace{3mm}Biot energy \ref{Biot-energy}
					\vspace{2mm}
				 \end{minipage}&$\times$&$\checkmark\checkmark$&$\times$&$\times$&$\checkmark\checkmark$&$\times$&$\times$&$\checkmark\checkmark$&$\times$&$\times$&$\times$&$\times$
				\\\hline \begin{minipage}{3.1cm}
					\footnotesize	
					\vspace{3mm}slightly Biot energy which does not explode \ref{Biotsde}
					\vspace{2mm}
				\end{minipage}&$\times$ &$\times$ &$\times$ & $\times$& $\times$&$\times$ &$\times$ &$\times$ &$\times$ &$\times$ &$\times$& $\times$
				\\\hline \begin{minipage}{3.1cm}
				\footnotesize	
				\vspace{3mm}slightly Biot energy which  explodes \ref{Biotse}
				\vspace{2mm}
			\end{minipage}&$\times$ &$\times$ &$\times$ & $\times$& $\times$&$\times$ &$\times$ &$(\times)$ &$\times$ &$\times$ &$\times$& ($\times$)
				\\\hline \begin{minipage}{3.1cm}
					\footnotesize	
					\vspace{3mm}Saint-Venant-Kirchhoff\\energy\ \  \ref{SV-energy}
					\vspace{2mm}
				\end{minipage}&$\times$&$\times$&$\times$&$\times$&$\times$&$\times$&$\times$&$\times$&$\times$&$\times$&$\times$&$\times$
				\\\hline \begin{minipage}{3cm}
					\footnotesize
					\vspace*{2mm}	slightly Mooney-Rivlin vol-iso split which explodes\ \ \ref{MRsc-energy}
					\vspace*{2mm}
				 \end{minipage}&$((\checkmark))$&$\times$&$((\checkmark))$&$((\checkmark))$&$\times$&$(\checkmark)$&$\times$&$\times$&$\times$&$\times$&$\times$&$(?)$
				\\\hline \begin{minipage}{3cm}
					\footnotesize
					\vspace*{2mm}	  the slightly exponentiated Mooney-Rivlin which does not explodes\ \ \ref{MRsc-energy0}
					\vspace*{2mm}
				\end{minipage}&$((\checkmark))$&$\times$&$\times$&$((\checkmark))$&$\times$&$\times$&$\times$&$\times$&$\times$&$\times$&$\times$&$\times$
				\\\hline
			 \begin{minipage}{3cm}
					\footnotesize
					\vspace*{2mm}	  the slightly exponentiated Mooney-Rivlin \ref{expslightlyMooney-energy}
					\vspace*{2mm}
				 \end{minipage}&$((\checkmark))$&$\times$&$((\checkmark))$&$((\checkmark))$&$\times$&$((\checkmark))$&$\times$&$\times$&$\times$&$\times$&$\times$&$(?)$
					\\\hline
				\begin{minipage}{3cm}
					\footnotesize
					\vspace*{2mm}	 $\frac{\|F\|^2}{(\det F)^{2/3}}$
					\vspace*{2mm}
				 \end{minipage}&$\checkmark$&$\times$&$\checkmark$&$\checkmark$&$\times$&$\checkmark$&$\times$&$\times$&$\times$&$\times$&$\times$&$\checkmark\checkmark$
				\\\hline
				\begin{minipage}{3cm}
					\footnotesize
					\vspace*{2mm}	 $\left\|\Cof \frac{F}{(\det F)^{1/3}}\right\|^2$
					\vspace*{2mm}
				 \end{minipage}&$\checkmark$&$\times$&$\checkmark$&$\checkmark$&$\times$&$\checkmark$&$\times$&$\times$&$\times$&$\times$&$\times$&$(\checkmark)$
				\\\hline \begin{minipage}{3cm}
					\footnotesize
					\vspace*{2mm}the 2-parameter Mooney-Rivlin material\ \   \ref{2parMR}
					\vspace*{2mm}
				\end{minipage}&$(\checkmark)$  &$\times$ &$(\checkmark)$ &$(\checkmark)$ &$\times$ &$(\checkmark)$ & $\times$& $\times$& $\times$& $\times$&$\times$& $(\checkmark)$
				\\\hline \begin{minipage}{3cm}
					\footnotesize\vspace*{2mm}	 slightly compressible Neo-Hooke not explode for $\kappa=\mu$\,\,\,\ref{NHsc-energy}
					\vspace*{2mm}
				\end{minipage}&$(\checkmark)$&$\times$&$\times$&$(\checkmark)$&$\times$&$\times$&$\times$&$\times$&$\times$&$\times$&$\times$&$\times\times$
					\\\hline \begin{minipage}{3cm}
					\footnotesize\vspace*{2mm}	 slightly compressible Neo-Hooke not explode for $\kappa=100\mu$\,\,\,\ref{NHsc-energy}
					\vspace*{2mm}
				\end{minipage}&$(\checkmark)$&$\times$&$\times$&$(\checkmark)$&$\times$&$\times$&$\times$&$\times$&$\times$&$\times$&$\times$&$\times\times$
					\\\hline \begin{minipage}{3.2cm}
					\footnotesize\vspace*{2mm}	 slightly compressible Neo-Hooke not explode for $\kappa=500\mu$\,\,\,\ref{NHsc-energy}
					\vspace*{2mm}
				\end{minipage}&$(\checkmark)$&$\times$&$\times$&$(\checkmark)$&$\times$&$\times$&$\times$&$\times$&$\times$&$\times$&$\times$&$\times\times$
				\\\hline \begin{minipage}{3cm}
					\footnotesize\vspace*{2mm}	 slightly compressible Neo-Hooke explode exponential $\kappa=\mu$\,\,\, \ref{NHvi-energy}
					\vspace*{2mm}
				 \end{minipage}&$(\checkmark)$&$(\times)$&$(\checkmark)$&$(\checkmark)$&$(\times)$&$(\checkmark)$&$\times$&$\times$&$\times$&$\times$&$\times$&$\checkmark\checkmark$
					\\\hline \begin{minipage}{3.2cm}
					\footnotesize\vspace*{2mm}	 slightly compressible Neo-Hooke explode exponential $\kappa=50\mu$\,\,\,\ref{NHvi-energy}
					\vspace*{2mm}
				 \end{minipage}&$(\checkmark)$&$(\times)$&$(\checkmark)$&$(\checkmark)$&$(\times)$&$(\checkmark)$&$\times$&$\times$&$\times$&$(\times)$&$\times$&$\checkmark\checkmark$
				\\\hline \begin{minipage}{3.2cm}
					\footnotesize\vspace*{2mm}	 slightly compressible Neo-Hooke explode exponential $\kappa=100\mu$\,\,\,\ref{NHvi-energy}
					\vspace*{2mm}
				 \end{minipage}&$(\checkmark)$&$\times$&$(\checkmark)$&$(\checkmark)$&$\times$&$(\checkmark)$&$\times$&$\times$&$\times$&$\times$&$\times$&$\checkmark\checkmark$
					\\\hline \begin{minipage}{3.2cm}
					\footnotesize\vspace*{2mm}	 slightly compressible Neo-Hooke explode exponential $\kappa=500\mu$\,\,\,\ref{NHvi-energy}
					\vspace*{2mm}
				 \end{minipage}&$(\checkmark)$&$\times$&$(\checkmark)$&$(\checkmark)$&$\times$&$(\checkmark)$&$\times$&$\times$&$\times$&$\times$&$\times$&$\checkmark\checkmark$
				\\
				\hline
		\end{tabular}}
		\caption{\footnotesize The Biot energy,  the Saint-Venant-Kirchhoff energy, Mooney-Rivlin type energies,  Neo-Hooke type energies}\label{tabelbc1}
	\end{table}
	
\end{footnotesize}

\begin{footnotesize}
	\begin{table}[hbt!]
		\centering
		\resizebox{16.5cm}{!}{ 	\begin{tabular} [hbt!]{|l|c|c|c|c|c|c|c|c|c|c|c|c|}
				\hline
				\footnotesize	\begin{minipage}{3cm}\footnotesize \vspace{3mm}\textbf{Energy}\vspace{3mm}\end{minipage}& \begin{minipage}{1cm} \footnotesize\textbf{$\sigma$-$V$ inv} \end{minipage}&\begin{minipage}{1.2cm} \footnotesize\textbf{$T_{\rm Biot}$-$U$ inv} \end{minipage} &\begin{minipage}{1cm} \footnotesize\textbf{$\tau$-$V$ inv} \end{minipage}& \begin{minipage}{1cm} \footnotesize\textbf{$\sigma$-$\log V$ inv} \end{minipage}&\begin{minipage}{1.2cm} \footnotesize\vspace{3mm}\textbf{$T_{\rm Biot}$-$\log U$ inv} \vspace{3mm}\end{minipage} &\begin{minipage}{1cm} \footnotesize\textbf{$\tau$-$\log V$ inv.} \end{minipage}
				& \begin{minipage}{1cm} \footnotesize\textbf{$\sigma$-$V$ mon} \end{minipage}&\begin{minipage}{1.2cm} \footnotesize\textbf{$T_{\rm Biot}$-$U$ mon} \end{minipage} &\begin{minipage}{1cm} \footnotesize\textbf{$\tau$-$V$ mon} \end{minipage}& \begin{minipage}{2cm} \footnotesize\textbf{$\sigma$-$\log V$ mon \\ (TSTS-M$^{++}$)} \end{minipage}&\begin{minipage}{1.2cm} \footnotesize\textbf{$T_{\rm Biot}$ -$\log U$ mon} \end{minipage} &\begin{minipage}{1.5cm} \footnotesize\textbf{$\tau$-$\log V$ mon\\ Hill's inequality} \end{minipage}
				\\\hline \begin{minipage}{3cm}
					\footnotesize\vspace*{2mm}	 original Ciarlet-Geymonat energy for compressible materials \ref{CG-energy}
					\vspace*{2mm}
				\end{minipage}&$(\checkmark)$&$\times$&$\checkmark$&$(\checkmark)$&$\times$&$\checkmark$&$\times$&$\times$&$\times$&$\times$&$\times$&$\checkmark$
				\\\hline \begin{minipage}{3cm}
					\footnotesize\vspace*{2mm}	 simplified  Ciarlet-Geymonat energy for compressible materials \ref{CG-energys}
					\vspace*{2mm}
				\end{minipage}&$\times$&$(\checkmark)$&$\checkmark$&$\times$&$(\checkmark)$&$\checkmark$&$\times$&$(\checkmark)$&$\checkmark$&$\times$&$(\checkmark)$&$\checkmark$
				\\
				\hline \begin{minipage}{3cm}
					\footnotesize\vspace*{2mm}	 Simo-Pister energy for compressible materials \ref{SP}
					\vspace*{2mm}
				\end{minipage}&$\times$&$\times$&$(\checkmark)$&$\times$&$\times$&$\checkmark$&$\times$&$\times$&$\times$&$\times$&$\times$&$\checkmark\checkmark$
				\\\hline \begin{minipage}{3cm}
					\footnotesize\vspace*{2mm}	 modified Simo-Pister energy for compressible materials \ref{mSP}
					\vspace*{2mm}
				\end{minipage}&$(?)$&$\times$&$((\checkmark))$&$(?)$&$\times$&$((\checkmark))$&$\times$&$\times$&$\times$&$\times$&$\times$&$((\checkmark))$	\\\hline \begin{minipage}{3cm}
					\footnotesize\vspace*{2mm}	 slightly compressible Ogden energy \ref{Ogdenvi-energy} for positive $\alpha_k$
					\vspace*{2mm}
				\end{minipage}&$(\checkmark)$&$\times$&$(\checkmark)$&$(\checkmark)$&$\times$&$(\checkmark)$&$\times$&$\times$&$\times$&$\times$&$\times$&$\checkmark\checkmark$
				\\\hline \begin{minipage}{3cm}
					\footnotesize\vspace*{2mm}	 slightly compressible Ogden energy for negative $\alpha_k$ \ref{Ogdenvi-energy}
					\vspace*{2mm}
				\end{minipage}&$((\checkmark))$&$\times$&$((\checkmark))$&$((\checkmark))$&$\times$&$((\checkmark))$&$\times$&$\times$&$\times$&$\times$&$\times$&$(?)$
				\\\hline \begin{minipage}{3cm}
					\footnotesize\vspace*{2mm}	  Fung-type energy  for slightly compressible materials \ref{Fuvi-energy}
					\vspace*{2mm}
				\end{minipage}&$(?)$&$(\times)$&$(?)$&$(?)$&$(\times)$&$(?)$&$\times$&$(\times)$&$\times$&$\times$&$\times$&$(?)$
				\\\hline \begin{minipage}{3cm}
					\footnotesize\vspace*{2mm}	 exponentiation of the Hencky-energy \ref{expHencky-energy}
					\vspace*{2mm}
				 \end{minipage}&$(\checkmark)$&$\times$&$(\checkmark)$&$(\checkmark)$&$\times$&$(\checkmark)$&$\times$&$\times$&$\times$&$\checkmark\checkmark$&$\times$&$\checkmark\checkmark$
				\\\hline\begin{minipage}{3cm}
					\footnotesize\vspace*{2mm}	  exponentiation of the isochoric Hencky-energy \ref{hyperHookei-energy}
					\vspace*{2mm}
				 \end{minipage}&$\checkmark$&$\times$&$\checkmark$&$\checkmark$&$\times$&$\checkmark$&$\times$&$\times$&$\times$&$\times$&$\times$&$\checkmark\checkmark$
				\\\hline \begin{minipage}{3cm}
					\footnotesize\vspace*{2mm}	  simplified exponentiation of the Hencky energy \ref{shyperHooke-energy}					 \vspace*{2mm}
				 \end{minipage}&$(\checkmark)$&$\times$&$(\checkmark)$&$(\checkmark)$&$\times$&$(\checkmark)$&$\times$&$\times$&$\times$&$\checkmark\checkmark$&$\times$&$\checkmark\checkmark$
				\\
				\hline
			 \begin{minipage}{3cm}
\footnotesize\vspace*{2mm}	  hyper Hencky:   iso-vol split exp-Hencky \ref{expHenckyvi-energy}
\vspace*{2mm}
\end{minipage}&$\checkmark\checkmark$&$\times$&$\checkmark\checkmark$&$\checkmark\checkmark$&$\times$&$\checkmark\checkmark$&$\times$&$\times$&$\times$&$\times$&$\times$&$\checkmark\checkmark$
\\\hline	\begin{minipage}{3cm}
	\footnotesize\vspace*{2mm}non	 iso-vol split exp-Hencky \ref{expHenckynvi-energy}
	\vspace*{2mm}
\end{minipage} &$((\checkmark))$&$\times$&$((\checkmark))$&$((\checkmark))$&$\times$&$((\checkmark))$&$\times$&$\times$&$\times$&$((\checkmark))$&$\times$&$((\checkmark))$

				\\\hline \begin{minipage}{3cm}
					\footnotesize\vspace*{2mm}	 Bazant energy \ref{Bazant-energy}
					\vspace*{2mm}
				\end{minipage}&$\times$&$\times$&$\times$&$\times$&$\times$&$\times$&$\times$&$\times$&$\times$&$\times$&$\times$&$\times$
				\\
				\hline
				\footnotesize	 \begin{minipage}{3cm}
					\footnotesize	
					\vspace*{2mm}
					slightly compressible Bazant \ref{Bazantsc-energy}
					\vspace*{2mm}
				\end{minipage}&$(?)$&$(\times)$&$(?)$&$(?)$&$(\times)$&$(?)$&$\times$&$\times$&$\times$&$\times$&$\times$&$(?)$
				\\\hline \begin{minipage}{3cm}
					\footnotesize	\vspace*{2mm}
					exponentiated Bazant \  \ \ref{expBazant-energy}
					\vspace*{2mm}
				\end{minipage}&$(?)$&$(\times)$&$(?)$&$(?)$&$(\times)$&$(?)$&$\times$&$\times$&$\times$&$(?)$&$\times$&$(?)$	
				\\\hline \begin{minipage}{3cm}
					\footnotesize	\vspace*{2mm}  exponentiated  slightly Bazant which does not explode\ \ \ref{expBazantsc-energyede}
					\vspace*{2mm}
				\end{minipage}&$(?)$&$\times$&$\times$&$(?)$&$\times$&$\times$&$\times$&$\times$&$\times$&$\times$&$\times$&$\times$	
				\\\hline \begin{minipage}{3cm}
					\footnotesize	\vspace*{2mm}  exponentiated  slightly Bazant which explodes\ \ \ref{expBazantsc-energye}
					\vspace*{2mm}
				\end{minipage}&$(?)$&$\times$&$(?)$&$(?)$&$\times$&$(?)$&$\times$&$\times$&$\times$&$\times$&$\times$&$(?)$
				\\
				\hline
				\begin{minipage}{3cm}
					\footnotesize\vspace*{2mm}	  Saravan energy \ref{Saravan-energy}
					\vspace*{2mm}
				\end{minipage}&$\times$&$\times$&$\times$&$\times$&$\times$&$\times$&$\times$&$\times$&$\times$&$\times$&$\times$&$\times$
				\\\hline
				\begin{minipage}{3cm}
					\footnotesize	\vspace*{2mm}  two parameter Signorini model \ref{Signorini}
					\vspace*{2mm}
				\end{minipage}&$\times$&$\times$&$\times$&$\times$&$\times$&$\times$&$\times$&$\times$&$\times$&$\times$&$\times$&$\times$\\
				\hline
				\begin{minipage}{3cm}
					\footnotesize	\vspace*{2mm} Gent energy  slightly compressible \ref{Gentsc-energy2}
					\vspace*{2mm}
				\end{minipage}&$\times$&$\times$&$(\checkmark)$&$\times$&$\times$&$(\checkmark)$&$\times$&$\times$&$\times$&$\times$&$\times$&$(\checkmark)$\\
				\hline
				\begin{minipage}{3cm}
				\footnotesize	\vspace*{2mm} Wollner's energy  \ref{Gte}
				\vspace*{2mm}
				\end{minipage}&$\checkmark\checkmark$&$\checkmark$ &$\checkmark$ & $\checkmark\checkmark$&$\checkmark$ &$\checkmark$ & $\times$&$\checkmark$ &$\times$ & $\checkmark\checkmark$ &$(\times)$&$\checkmark$\\\hline	\begin{minipage}{3cm}
				\footnotesize\vspace*{2mm}exp-Hencky with limited compressibility \ref{expHencky-tan}
				\vspace*{2mm}
				\end{minipage} &$\checkmark\checkmark$&$\times$&$\checkmark$&$\checkmark\checkmark$&$\times$&$\checkmark$&$\times$&$\times$&$\times$&$\checkmark\checkmark$&$\times$&$\checkmark$
				\\\hline \begin{minipage}{3cm}
				\footnotesize\vspace*{2mm}	 slightly compressible Benam  energy which does not explode  \ref{Benam-energy}
				\vspace*{2mm}
				\end{minipage}&$(\times)$&$\times$&$\times$&$(\times)$&$\times$&$\times$&$\times$&$\times$&$\times$&$\times$&$\times$&$\times$
				\\\hline \begin{minipage}{3cm}
				\footnotesize\vspace*{2mm}	 slightly compressible Benam  energy which  explodes \ \  \ref{Benam-energye}
				\vspace*{2mm}
				\end{minipage}&$(\times)$&$\times$&$(\times)$&$(\times)$&$\times$&$(\times)$&$\times$&$\times$&$\times$&$(\times)$&$\times$&$(\times)$ \\
				\hline
		\end{tabular}}
		\caption{\footnotesize   Ciarlet-Geymonat-type energies, Hencky-type energies, Bazant-type energies, Bazant-type energies, Signorini energy, Gent energy, Benam-type energy, Wollner energy}\label{tabelbc2Capta}
	\end{table}
	
\end{footnotesize}

\begin{footnotesize}
	\begin{table}[hbt!]
		\centering
		\resizebox{16.5cm}{!}{ 	\begin{tabular} [hbt!]{|l|c|c|c|c|c|c|c|c|c|c|c|c|}
				\hline
				\footnotesize	\begin{minipage}{3cm}\footnotesize \vspace{3mm}\textbf{Energy}\vspace{3mm}\end{minipage}& \begin{minipage}{1cm} \footnotesize\textbf{$\sigma$-$V$ inv} \end{minipage}&\begin{minipage}{1.2cm} \footnotesize\textbf{$T_{\rm Biot}$-$U$ inv} \end{minipage} &\begin{minipage}{1cm} \footnotesize\textbf{$\tau$-$V$ inv} \end{minipage}& \begin{minipage}{1cm} \footnotesize\textbf{$\sigma$-$\log V$ inv} \end{minipage}&\begin{minipage}{1.2cm} \footnotesize\vspace{3mm}\textbf{$T_{\rm Biot}$-$\log U$ inv} \vspace{3mm}\end{minipage} &\begin{minipage}{1cm} \footnotesize\textbf{$\tau$-$\log V$ inv.} \end{minipage}
				& \begin{minipage}{1cm} \footnotesize\textbf{$\sigma$-$V$ mon} \end{minipage}&\begin{minipage}{1.2cm} \footnotesize\textbf{$T_{\rm Biot}$-$U$ mon} \end{minipage} &\begin{minipage}{1cm} \footnotesize\textbf{$\tau$-$V$ mon} \end{minipage}& \begin{minipage}{2cm} \footnotesize\textbf{$\sigma$-$\log V$ mon \\ (TSTS-M$^{++}$)} \end{minipage}&\begin{minipage}{1.2cm} \footnotesize\textbf{$T_{\rm Biot}$ -$\log U$ mon} \end{minipage} &\begin{minipage}{1.5cm} \footnotesize\textbf{$\tau$-$\log V$ mon\\ Hill's inequality} \end{minipage}
				\\\hline  
				\begin{minipage}{3cm}
					\footnotesize	\vspace*{2mm}  tension-compression symmetric energy \ref{tc-energy}
					\vspace*{2mm}
				\end{minipage}&$(?)$&$\times$&$(?)$&$(?)$&$\times$&$(?)$&$\times$&$\times$&$\times$&$\times$&$\times$&$(?)$\\
				\hline
				\begin{minipage}{3cm}
					\footnotesize	\vspace*{2mm}  exponentiated tension-compression symmetric energy\ \   \ref{exptc-energy}
					\vspace*{2mm}
				\end{minipage}&$(?)$&$\times$&$(?)$&$(?)$&$\times$&$(?)$&$\times$&$\times$&$\times$&$\times$&$\times$&$(?)$\\
				\hline
				\begin{minipage}{3cm}
					\footnotesize	\vspace*{2mm} Richter energy \ref{Richter-energy}
					\vspace*{2mm}
				\end{minipage}&$\checkmark$&$\times$&$\times$&$\checkmark$&$\times$&$\times$&$\checkmark$&$\times$&$\times$&$\times$&$\times$&$\times$
				\\\hline
				\begin{minipage}{3cm}
					\footnotesize\vspace*{2mm}	energy  convex   in $C$ \ref{convex-energy}
					\vspace*{2mm}
				\end{minipage}&$\times$&$(\checkmark)$&$\checkmark$&$\times$&$(\checkmark)$&$\checkmark$&$\times$&$(\checkmark)$&$\checkmark$&$\times$&$(\checkmark)$&$\checkmark$\\
				\hline
				\begin{minipage}{3cm}
					\footnotesize\vspace*{2mm}	 slightly compressible Carroll  energy \ref{Carolvi-energy}
					\vspace*{2mm}
				\end{minipage}&$(\times)$&$\times$&$(\times)$&$(\times)$&$\times$&$(\times)$&$\times$&$\times$&$\times$&$\times$&$\times$&$(\times)$\\
				\hline
				\begin{minipage}{3cm}
					\footnotesize\vspace*{2mm}	 modified slightly compressible Carroll  energy \ref{mCarolvi-energy}
					\vspace*{2mm}
				\end{minipage}&$(\times)$&$\times$&$(\times)$&$(\times)$&$\times$&$(\times)$&$\times$&$\times$&$\times$&$\times$&$\times$&$(\times)$\\
				\hline
				\begin{minipage}{3cm}
					\footnotesize\vspace*{2mm}	 Gao  energy \ref{Carolvi-energy}
					\vspace*{2mm}
				\end{minipage}&$(?)$&$(?)$&$(?)$&$(?)$&$(?)$&$(?)$&$\times$&$(?)$&$(?)$&$\times$&$(?)$&$(?)$\\
				\hline
				\begin{minipage}{3cm}
					\footnotesize\vspace*{2mm}	 Murnaghan energy \ref{Carolvi-energy}
					\vspace*{2mm}
				\end{minipage}&$\times$&$\times$&$\times$&$\times$&$\times$&$\times$&$\times$&$\times$&$\times$&$\times$&$\times$&$\times$\\
				\hline
		\end{tabular}}
		\caption{\footnotesize   Other-type  of elastic energies.}\label{tabelbc3Capta}
	\end{table}
	
\end{footnotesize}

\begin{footnotesize}
	\begin{table} [hbt!]
		\centering
		\resizebox{16.5cm}{!}{ 	\begin{tabular} [hbt!]{|l|c|c|c|c|c|c|c|c|c|c|c|c|}
				\hline
				\footnotesize	\begin{minipage}{3cm}\footnotesize \textbf{Cauchy stress tensor}\end{minipage}& \begin{minipage}{1cm} \footnotesize\textbf{$\sigma$-$V$ inv} \end{minipage}&\begin{minipage}{1.2cm} \footnotesize\textbf{$T_{\rm Biot}$-$U$ inv} \end{minipage} &\begin{minipage}{1cm} \footnotesize\textbf{$\tau$-$V$ inv} \end{minipage}& \begin{minipage}{1cm} \footnotesize\textbf{$\sigma$-$\log V$ inv} \end{minipage}&\begin{minipage}{1.2cm} \footnotesize\vspace*{2mm}\textbf{$T_{\rm Biot}$-$\log U$ inv}\vspace*{2mm} \end{minipage} &\begin{minipage}{1cm} \footnotesize\textbf{$\tau$-$\log V$ inv.} \end{minipage}
				& \begin{minipage}{1cm} \footnotesize\textbf{$\sigma$-$V$ mon} \end{minipage}&\begin{minipage}{1.2cm} \footnotesize\textbf{$T_{\rm Biot}$-$U$ mon} \end{minipage} &\begin{minipage}{1cm} \footnotesize\textbf{$\tau$-$V$ mon} \end{minipage}& \begin{minipage}{2cm} \footnotesize\textbf{$\sigma$-$\log V$ mon \\ (TSTS-M$^{++}$)} \end{minipage}&\begin{minipage}{1.2cm} \footnotesize\textbf{$T_{\rm Biot}$ -$\log U$ mon} \end{minipage} &\begin{minipage}{1.5cm} \footnotesize\textbf{$\tau$-$\log V$ mon\\ Hill's inequality} \end{minipage}
				\\\hline \begin{minipage}{3cm}
					\footnotesize	\vspace*{2mm}
					$\sigma_{\rm NH}(B)=$\\$\frac{1}{\sqrt{\det(B)}}
					\, (B-\id)$\vspace*{2mm}
				 \end{minipage}&$(\times)$&$(\checkmark)$&$\checkmark$&$(\times)$&$(\checkmark)$&$(\checkmark)$&$\times$&$(\checkmark)$&$(\checkmark)$&$(\times)$&$(\checkmark)$&$(\checkmark)$\\
				\hline \begin{minipage}{3cm}
					\footnotesize\vspace*{2mm}
					$\sigma(B)=B-B^{-1}$\vspace*{2mm}
				 \end{minipage}&$\checkmark\checkmark$&$((\checkmark))$&$\times$&$\checkmark\checkmark$&$(\checkmark)$&$\times$&$((\checkmark))$&$\times$&$\times$&$\checkmark\checkmark$&$\times$&$\times$\\
				\hline
				\begin{minipage}{3cm}
					\footnotesize\vspace*{2mm}
					$\sigma(B)=B-\id$\vspace*{2mm}
				 \end{minipage}&$\checkmark$&$\times$&$\times$&$\checkmark$&$\times$&$\times$&$\checkmark$&$\times$&$\times$&$\checkmark\checkmark$&$\times$&$\times$\\
				\hline
				\begin{minipage}{3cm}
					\footnotesize\vspace*{2mm}
					$\sigma(B)=$\\ $B-\id+\tr(B-\id)\, \id$\vspace*{2mm}
				\end{minipage}&$\checkmark$&$\times$&$\times$&$\checkmark$&$\times$&$\times$&$\times$&$\times$&$\times$&$\times$&$\times$&$\times$\\
				\hline
				\begin{minipage}{3cm}
					\footnotesize\vspace*{2mm}
					$\sigma_{\rm Richter}(V)=$\\$V-\id+\tr(V-\id)\, \id$\vspace*{2mm}
				\end{minipage}&$\checkmark$&$\times$&$\times$&$\checkmark$&$\times$&$\times$&$\checkmark$&$\times$&$\times$&$\times$&$\times$&$\times$\\
				\hline
				\begin{minipage}{3cm}
					\footnotesize\vspace*{2mm}
					$\sigma_{\rm Hencky}(B)=\log B$\vspace*{2mm}
				 \end{minipage}&$\checkmark\checkmark$&$\times$&$\times$&$\checkmark\checkmark$&$\times$&$\times$&$\checkmark$&$\times$&$\times$&$\checkmark\checkmark$&$\times$&$\times$\\
				\hline
				\begin{minipage}{3cm}
					\footnotesize\vspace*{2mm}
					$\sigma_{\rm Blatz-Ko}(B)=$\\$\frac{1}{\det B}(\sqrt{\det B}\, B-\id)$\vspace*{2mm}
				 \end{minipage}&$\times$&$((\checkmark))$&$(\checkmark)$&$\times$&$(\checkmark)$&$(\checkmark)$&$\times$&$((\checkmark))$&$\times$&$\times$&$\times$&$(\checkmark)$
				\\\hline
				\begin{minipage}{3cm}
					\footnotesize\vspace*{2mm}
					$T_{\rm Biot}^{\rm  Becker}(U)=\log U$\vspace*{2mm}
				 \end{minipage}&$\times$&$\checkmark\checkmark$&$\times$&$\times$&$\checkmark\checkmark$&$\times$&$\times$&$\checkmark\checkmark$&$\times$&$\times$&$\checkmark\checkmark$&$\times$
					\\\hline
				\begin{minipage}{3cm}
					\footnotesize\vspace*{2mm}
					$\sigma(B)=\frac{\mu}{4}(B-B^{-1})$\\\hspace*{0.9cm} $+\frac{\lambda}{4}\log \det B \cdot \id.$\vspace*{2mm}
				\end{minipage}&$\checkmark\checkmark$ &$\times$ &$\times$ &$\checkmark\checkmark$ &$\times$ &$\times$ &$\checkmark\checkmark$ &$\times$ & $\times$&$\checkmark\checkmark$ &$\times$ &$\times$
					\\\hline
				\begin{minipage}{3.3cm}
					\footnotesize\vspace*{2mm}
					$\sigma(B)=\frac{\mu}{4}(\frac{B}{\det B^{1/3}}-\id)$\\\hspace*{0.9cm} $+\frac{\lambda}{4}\log \det B \cdot \id.$\vspace*{2mm}
				\end{minipage}&$\checkmark$&$\times$&$\times$&$\checkmark$&$\times$&$\times$&$\times$&$\times$&$\times$&$\times$&$\times$&$\times$ \\
				\hline
		\end{tabular}}
		\caption{\footnotesize Cauchy-elasticity: some direct Cauchy stress-strain pairs. Not hyperelastic, in general.}\label{tabelbc3}
	\end{table}
	
\end{footnotesize}

\newpage

\clearpage
\begin{footnotesize}

	\bibliographystyle{plain} %plain

\appendix

\section{Appendix}
\subsection{General notation}\setcounter{equation}{0}

For $a,b\in\R^n$ we let $\langle {a},{b}\rangle_{\R^n}$  denote the scalar product on $\R^n$ with the
associated vector norm $\|a\|_{\R^n}^2=\langle {a},{a}\rangle_{\R^n}$.
We denote by $\mathbb{R}^{n\times n}$, $n\in\mathbb{N}$, the set of real $n\times n$ second order tensors, written in
capital letters. We adopt the usual abbreviations of Lie-group theory, i.e.,
${\rm GL}(n)=\{X\in\mathbb{R}^{n\times n}\;|\det({X})\neq 0\}$ the general linear group,
${\rm SL}(n)=\{X\in {\rm GL}(n)\;|\det({X})=1\},\;
\mathrm{O}(n)=\{X\in {\rm GL}(n)\;|\;X^TX=\id_n\},\;{\rm SO}(n)=\{X\in {\rm GL}(n)| X^TX=\id_n,\det({X})=1\}$ with
corresponding Lie-algebras $\mathfrak{so}(n)=\{X\in\mathbb{R}^{n\times n}\;|X^T=-X\}$ of skew symmetric tensors
and $\mathfrak{sl}(n)=\{X\in\mathbb{R}^{n\times n}\;| \tr({X})=0\}$ of traceless tensors.  {Here and i}n the following the superscript
$^T$ is used to denote transposition.
The standard Euclidean scalar product on  the set of real $n\times  {m}$ second order tensors $\mathbb{R}^{n\times  {m}}$ is given by
$\bigl\langle  {X},{Y} \bigr\rangle _{\mathbb{R}^{n\times  {m}}}={\rm tr}(X\, Y^T)$, and thus the  {(squared)} Frobenius tensor norm is
$\lVert {X}\rVert ^2_{\mathbb{R}^{n\times  {m}}}=\bigl\langle  {X},{X} \bigr\rangle _{\mathbb{R}^{n\times  {m}}}$.  In the following we omit the index
$\mathbb{R}^n,\mathbb{R}^{n\times m}$. The identity tensor on $\mathbb{R}^{n \times n}$ will be denoted by $\id_n$, so that
${\rm tr}({X})= \bigl\langle {X},{\id}_n\bigr\rangle $.  We let ${\rm Sym}(n)$ and ${\rm Sym}^{++}(n)$ denote the symmetric and positive definite symmetric tensors, respectively.  For all $X\in\mathbb{R}^{3\times3}$ we set ${\rm sym}\, X=\frac{1}{2}(X^T+X)\in{\rm Sym}(3)$, $\skw\, X\,=\frac{1}{2}(X-X^T)\in \mathfrak{so}(3)$ and the deviatoric part $\dev  \, X=X-\frac{1}{n}\;\tr(X)\,\id_n\in \mathfrak{sl}(n)$  and we have
the \emph{orthogonal Cartan-decomposition  of the Lie-algebra} $\mathfrak{gl}(3)$
\begin{align*}
	\mathfrak{gl}(3)&=\{\mathfrak{sl}(3)\cap {\rm Sym}(3)\}\oplus\mathfrak{so}(3) \oplus\mathbb{R}\!\cdot\! \id_3,\qquad\qquad
	X=\dev  \,\sym \,X+ \skw\,X+\frac{1}{3}\tr(X)\, \id_3\,.
\end{align*}

For $X\in {\rm GL}(n)$, $\Cof X = (\det X)X^{-T}$ is the cofactor of $X\in {\rm GL}(n)$, while  ${\rm Adj}({X})$  denotes the tensor of
transposed cofactors.
For vectors $\xi,\eta\in\mathbb{R}^n$, we have the tensor product
$(\xi\otimes\eta)_{ij}=\xi_i\,\eta_j$.

For vectors $v=\left(v_1,v_2,v_3\right)^T\ \in\R^3,
$ we define
$
{\rm diag}\, v=\left(
\begin{array}{ccc}
	v_1 & 0 & 0 \\
	0 & v_2 & 0 \\
	0 & 0 & v_3 \\
\end{array}
\right),
$
while for a matrix
$
F=\left(
\begin{array}{ccc}
	F_{11} & F_{12} & F_{13} \\
	F_{21} & F_{22} & F_{23} \\
	F_{31} & F_{32} & F_{33} \\
\end{array}
\right)\in\R^{3\times 3}
$ we let
$
{\rm vect}\, F=(F_{11}, F_{12}, F_{13}, F_{21}, F_{22},F_{23}, F_{31}, F_{32}, F_{33})^T\in \R^9.
$

The Fr\'echet derivative of a function $W:\mathbb{R}^{n\times n}\to \mathbb{R}$ at $F\in \mathbb{R}^{n\times n}$ applied to the tensor-valued increment $H$ is denoted by ${\rm D}_F[W(F)]. H$. Similarly, ${\rm D}_F^2[W(F)]. (H_1,H_2)$ is the bilinear form induced by the second Fr\'echet derivative of the function $W$ at $F$ applied to $(H_1,H_2)$.

We also identify the first derivative ${\rm D}_F W$ with the gradient, writing ${\rm D}_F W.H = \iprod{{\rm D}_F W, H}$ for $F\in\GLpn$ and $H\in\Rnn$, and employ the chain rules
\begin{align*}
	{\rm D}_F(( \Phi\circ W)(F)).H={\rm D}_F(\Phi(W(F))).H= \Phi'(W(F))\, {\rm D}_FW(F).H\,,\notag\\
	{\rm D}_F((W\circ G)(F)).H={\rm D}_F(W(G(F))).H= \langle DW(G(F)), {\rm D}_FG(F).H\rangle\notag
\end{align*}
for $W:\mathbb{R}^{3\times3}\to \mathbb{R}$, $G:\mathbb{R}^{3\times3}\to \mathbb{R}^{3\times3}$ and $\Phi:\mathbb{R}\to \mathbb{R}$.z

We also recall some useful identities:
\begin{itemize}
	\item $\tr(B^{-1}X\, B)=\tr(X)$ for any invertible matrix $B$.
	\item $\dev_n(B^{-1}X\, B)=B^{-1}X\, B-\frac{1}{n}\,\tr(B^{-1}X\, B)\,\id=B^{-1}(\dev_n X)\, B$ for any invertible matrix $B$.
	\item  $\norm{\dev_n X}^2=\norm{X-\frac1n\tr X \cdot \id}^2=\norm{X}^2+\frac1{n^2}(\tr X)^2\norm{\id}^2-\frac2n\tr X\langle X,I\rangle=\norm{X}^2-\frac1n(\tr
	X)^2$.
	\item The norm of the deviator in $\R^{n\times n}$:
	\begin{align*}
		\|\dev_n\left(
		\begin{array}{cccc}
			\xi_1&0&\cdots&0\\
			0&\xi_2&\cdots&0\\
			\vdots&\vdots &\ddots &\vdots\\
			0&0&\cdots&\xi_n\\
		\end{array}\right)
		\|^2&=\sum\limits_{i=1}^n \xi_i^2-\tel n(\sum\limits_{i=1}^n \xi_i)^2=\frac{1}{n}\sum\limits_{i,j=1,i<j}^n (\xi_i-\xi_j)^2.\notag
	\end{align*}
	\item $\log U=\sum\limits_{i=1}^n \log \lambda_i\,  N_i\otimes N_i,$ where $N_i$ are the eigenvectors of $U$ and $\lambda_i$ are the eigenvalues of
	$U$.
	
	\item $\log V=\sum\limits_{i=1}^n \log \widehat{\lambda}_i\,  \widehat{N}_i\otimes \widehat{N}_i,$ where $\widehat{N}_i$ are the eigenvectors of $V$
	and $\widehat{\lambda}_i$ are the eigenvalues of $V$.
\end{itemize}

For a fourth order linear mapping $\C : \Sym(3) \to \Sym(3)$ we agree on the following convention. \\
\\
We say that $\C$ has \emph{minor symmetry} if
\begin{align}
	\C.S \in \Sym(3) \qquad \forall \, S \in \Sym(3).
\end{align}
This can also be written in index notation as $C_{ijkm} = C_{jikm} = C_{ijmk}$. If we consider a more general fourth order tensor $\C : \R^{3 \times 3} \to \R^{3 \times 3}$ then $\C$ can be transformed having minor symmetry by considering the mapping $X \mapsto \sym(\C. \sym X)$ such that $\C: \R^{3 \times 3} \to \R^{3 \times 3}$ is minor symmetric, if and only if $\C.X = \sym(\C.\sym X)$. \\
\\
We say that $\C$ has \emph{major symmetry} (or is \emph{self-adjoint}, respectively) if
\begin{align}
	\langle \C. S_1, S_2 \rangle = \langle \C. S_2, S_1 \rangle \qquad \forall \, S_1, S_2 \in \Sym(3).
\end{align}
Major symmetry in index notation is understood as $C_{ijkm} = C_{kmij}$. \\
\\
The set of positive-definite, major symmetric fourth order tensors mapping $\R^{3 \times 3} \to \R^{3 \times 3}$ is denoted as $\Sym^{++}_4(9)$, in case of additional minor symmetry, i.e.~mapping $\Sym(3) \to \Sym(3)$ as $\Sym^{++}_4(6)$. By identifying $\Sym(3) \cong \R^6$, we can view $\C$ as a linear mapping in matrix form $\widetilde \C: \R^6 \to \R^6$. \newline If $H \in \Sym(3) \cong \R^6$ has the entries $H_{ij}$, we can write
\begin{align}
	\label{eqvec1}
	h = \text{vec}(H) = (H_{11}, H_{22}, H_{33}, H_{12}, H_{23}, H_{31}) \in \R^6 \qquad \text{so that} \qquad \langle \C.H, H \rangle_{\Sym(3)} = \langle \widetilde \C.h, h \rangle_{\R^6}.
\end{align}
If $\C: \Sym(3) \to \Sym(3)$, we can define $\sym\, \C$ by
\begin{align}
	\langle \C.H, H \rangle_{\Sym(3)} = \langle \widetilde \C.h, h \rangle_{\R^6} = \langle \sym \widetilde \C. h, h \rangle_{\R^6} =: \langle \sym \C.H, H \rangle_{\Sym(3)}, \qquad \forall \, H \in \Sym(3).
\end{align}
Major symmetry in these terms can be expressed as $\widetilde \C \in \Sym(6)$. \emph{In this text, however, we omit the tilde-operation and ${\bf sym}$ and write in short $\sym\C\in {\rm Sym}_4(6)$ if no confusion can arise.} In the same manner we speak about $\det \C$ meaning $\det \widetilde \C$. \\
\\
A linear mapping $\C : \R^{3 \times 3} \to \R^{3 \times 3}$ is positive definite if and only if
\begin{align}
	\label{eqposdef1}
	\langle \C.H, H \rangle > 0 \qquad \forall \, H \in \R^{3 \times 3} \qquad \iff \qquad \C \in \Sym^{++}_4(9)
\end{align}
and analogously it is positive semi-definite if and only if
\begin{align}
	\label{eqpossemidef1}
	\langle \C.H, H \rangle \ge 0 \qquad \forall \, H \in \R^{3 \times 3} \qquad \iff \qquad \C \in \Sym^+_4(9).
\end{align}
For $\C: \Sym(3) \to \Sym(3)$, after identifying $\Sym(3) \cong \R^6$, we can reformulate \eqref{eqposdef1} as $\widetilde \C \in \Sym^{++}(6)$ and \eqref{eqpossemidef1} as $\widetilde \C \in \Sym^+(6)$.

\subsection{Geometry of nonlinear  hyperelasticity }\label{appendixA}
Let $\Omega\subset{\R^n}$ ($n=1,2,3$)  be a bounded domain with Lipschitz boundary
$\partial\Omega$. A mapping $\varphi\col\Omega\to\R$ describes the deformation of the domain $\Omega$. The domain $\Omega$ is called the initial configuration (undeformed), while its image $\Omega_c:=\varphi(\Omega)$ is called the actual (deformed) configuration. Each of these configurations could be considered as referential configuration, depending on the practical problem we solve or model.  Another kinematic used variable, beside $\varphi$, is the displacement $u(x)=\varphi(x) - x$.

Since self-intersection of the material is not allowed,  there exists the inverse mapping $\varphi\col\Omega_c\to\Omega$ from the deformed configuration to its initial configuration. Therefore, the  \emph{deformation gradient} defined by
\begin{align*}
	F\colonequals{\rm D}\varphi\in \mathbb{R}^{n\times n}
\end{align*}
satisfies $F\in\GLp(n)$, i.e., $J=\det F>0$.

From a geometrical or analytical point of view, this would have been enough for a complete description of the deformation. However, in elasticity theory we consider  that the domain $\Omega$ is filled by an elastic body and the aim is to take into account (to model) the physical response of the body, i.e., the intrinsic relation between stress (internal forces) and strain (amount of deformation).

The tensor which describes the force of the deformed material per original (stress) is  the first Piola-Kirchhoff stress tensor, denoted by $S_1(F)$.
In the context of nonlinear hyperelasticity, where generalized convexity properties have an especially long and rich history \cite{Ball77,knowles1976failure,knowles1978failure}, the material behaviour of an elastic solid is described by a potential energy function
\begin{equation}
W\col\GLp(n)\to\R\,,\quad F\mapsto W(F)
\end{equation}
defined on the group $\GLp(n)$ of invertible matrices with positive determinants.
In hyperelasticity, the stress-strain relation is described by an energy density potential $W:\GLp(n)\to \mathbb{R}$, through $S_1={\rm D}_F[W(F)]$.  We assume that  the material is homogeneous. The energy density function $W$ represents the elastic energy measured per unit volume of the reference configuration instead of the energy of the whole body, defined as
\begin{align}
	I(\varphi)=\int_{\Omega} W({\rm D}\, \varphi(x))
	\, dx.	
\end{align}

Since in general it is not easy to work with tensors (matrices), in three dimensions, we consider the singular values (principal stretches) $\lambda_1$, $\lambda_2$, $\lambda_3$ of $F$, i.e. the eigenvalues  of  $U$, and the principal isotropic invariants of $U$
\begin{align}
	I_1&=\lambda_1+\lambda_2+\lambda_3=\tr(U)\,, \ \
	I_2=\lambda_1\lambda_2+\lambda_2\lambda_3+\lambda_3\lambda_1=\tr(\Cof U)\,,\ \
	I_3=\lambda_1\lambda_2\lambda_3=\det U\,.
\end{align}

Various tensors are considered as strain-tensors, e.g.,
\begin{itemize}
	\item
	$C=F^T F$ the right Cauchy-Green strain tensor;
	\item$B=F\, F^T$ the left Cauchy-Green (or Finger) strain tensor;
	\item$U$ the right
	stretch tensor, i.e., the unique element of ${\rm Sym}^{++}(n)$ for which $U^2=C$;
	\item$V$ the  left stretch tensor, i.e., the unique element of ${\rm
		Sym}^{++}(n)$ for which $V^2=B$,
	\item $\log U$ or $\log V$ the Hencky strain tensor, i.e.,
\end{itemize}
particular tensors from to the commonly used  \emph{strain tensors} of \emph{Seth-Hill type} \cite{seth1961generalized,hill1970constitutive}
\begin{align}
	E_{(m)}=\left\{
	\begin{array}{ll}
		\dd\frac{1}{2\, m}(U^{2\, m}-\id),& m\neq 0, \vspace{1.2mm}\\
		\log U,& m=0\,,
	\end{array}
	\right.
\end{align}
as well as various stress-tensors, e.g.,
\begin{itemize}
	\item  $S_1={\rm D}_F[{W}(F)]$
	the first  Piola-Kirchhoff stress tensor;
	\item  $S_2=F^{-1}S_1=2\,{\rm D}_C[\overline{W}(C)]$
	the second  Piola-Kirchhoff stress tensor (transforms force of the deformed material per original area back to the reference configuration);
	\item  $\sigma=\frac{1}{J}\,  S_1\, F^T=\frac{2}{J}\, {\rm D}_BW(B)\, B=\frac{1}{J}\,  {\rm D}_{\log V}\widehat{W}(\log V)\, B$ the Cauchy stress tensor, describes the stress of the deformed material with respect to areas in the deformed configuration $\varphi(\Omega)$, the ''true'' stress tensor;
	\item
	$
	T_{\rm Biot}=R^TS_1(F)={\rm D}_U[\widetilde{W}(U)]\,,
	$
	the Biot stress tensor,	which only uses the inverse rotation $R^T$ of the polar decomposition $F=R\,U=V\,R$ instead;
	\item  $\tau=J\, \sigma={\rm D}_{\log V}\widehat{W}(\log V)$ the Kirchhoff stress, widely used in numerical algorithms in metal plasticity (where there is no change in volume during plastic deformation).
\end{itemize}

Assumptions on the stress-strain relation are known as constitutive requirements.
Different materials vary in their elasticity behaviour. Since in  hyperelasticity the stress is defined by the energy density, the choice of  an  energy function is a constitutive choice.

 Moreover, many physical properties of the elastic material may be described in terms of the function $g$ and $\psi$, while $\Phi$ and $P$ are more suitable for mathematical purpose.

In many situations the stress-strain relations are characterized by the relation between their corresponding principal values, i.e., by the relations between $\lambda_1, \lambda_2, \lambda_3 $ and
\begin{itemize}
	\item  $\sigma_i =\dd\frac{1}{\lambda_1\lambda_2\lambda_3}\dd\lambda_i\frac{\partial g(\lambda_1,\lambda_2,\lambda_3)}{\partial \lambda_i}=\dd\frac{1}{\lambda_j\lambda_k}\dd\frac{\partial g(\lambda_1,\lambda_2,\lambda_3)}{\partial \lambda_i}, \ \ i\neq j\neq k \neq i$  \quad the principal Cauchy stresses, where $g:\mathbb{R}_+^3\to \mathbb{R}$ is the unique function  of the singular values of $U$ (principal stretch) such that $W(F)=\widetilde{W}(U)=g(\lambda_1,\lambda_2,\lambda_3)$,
	\item  $\sigma_i =\dd\frac{1}{\lambda_1\lambda_2\lambda_3}\frac{\partial \widetilde{g}(\log \lambda_1,\log \lambda_2,\log \lambda_3)}{\partial \log \lambda_i}$,\quad  where $\widetilde{g}:\mathbb{R}^3\to \mathbb{R}$ is the unique function  such that \\$\widetilde{g}(\log \lambda_1,\log \lambda_2,\log \lambda_3):=g(\lambda_1,\lambda_2,\lambda_3)$,
	\item 	  $\tau_i =J\, \sigma_i=\dd\lambda_i\frac{\partial g(\lambda_1,\lambda_2,\lambda_3)}{\partial \lambda_i}=\frac{\partial \widetilde{g}(\log \lambda_1,\log \lambda_2,\log \lambda_3)}{\partial \log \lambda_i},$
	\item   $T_i=\,\lambda_j\,\lambda_k\,\sigma_i= \dd\frac{\partial g(\lambda_1,\lambda_2,\lambda_3)}{\partial \lambda_i},\ i\neq j\neq k \neq i$ \quad the  principal forces (principal Biot-stresses).
\end{itemize}

In the modelling process of the equilibrium of elastic bodies it is  natural to assume that the material $\Omega$ always tends to minimize its elastic energy $I(\varphi)$.
Thus, once an energy is chosen, the deformation is given  by the solution of  the minimization problem
\begin{equation}
	I(\varphi)=\int_\Omega W({\rm D}\,\varphi(x))\,\dx\to\textrm{\ \ min.} \quad  {\rm   w.r.t. }\quad \varphi
\end{equation}
in an admissible set (in which we incorporate the boundary conditions, too).

However, there exists no mathematical model in classical nonlinear elasticity
\begin{itemize}
	\item  which is capable of describing the correct physical mechanical behaviour for any elastic material, especially for large strains
	\item[] \qquad \textbf{and}
	\item for which the existence of the minimizer of the corresponding variational problem is assured.
\end{itemize}

\subsection{Consistency with linear elasticity}\label{cle}
For isotropic hyperelastic materials to be consistent with isotropic linear elasticity, the stress-strain relation should have the following form in the infinitesimal strain limit
\begin{align}
	\sigma=2\, \mu\, \varepsilon+\lambda\, \tr(\varepsilon)\, \id,
\end{align}
where $\lambda$, $\mu$ are the Lam\'e parameters. The strain energy density function that corresponds to the above is
\begin{align}
	W=\mu\|\varepsilon\|^2+\frac{\lambda}{2} [\tr(\varepsilon)]^2.\end{align}
	
	For any nonlinear strain energy density function $W(\lambda_1,\lambda_2,\lambda_3)$ defined in the stretches $(\lambda_1,\lambda_2,\lambda_3)$ to reduce to the above forms for small strains, the following conditions have to be met
	\begin{align}
	&	W(1,1,1)=0, \qquad \qquad \frac{\partial W}{\partial \lambda_i}(1,1,1)=0,\notag\\
		&\left(\frac{\partial^2 W}{\partial \lambda_i\partial \lambda_j}(1,1,1)\right )_{i,j=1,2,3}=\lambda+2\,\mu\,\delta_{ij}=\begin{pmatrix}
			\lambda+2\mu&\lambda&\lambda\\
			\lambda&\lambda+2\mu&\lambda\\
			\lambda&\lambda&\lambda+2\mu
		\end{pmatrix},
	\end{align}
	see the book by Ogden \cite[p. 349]{Ogden83}. These relations are used to properly  identify the two isotropic material parameters $\mu$ and $\lambda$ at the identity in our list of energies.

\end{footnotesize}

\end{document}